\documentstyle[12pt,epsfig,myext]{article}
\begin{document}

\def\uncatcodespecials{\def\do##1{\catcode`##1=12 }\dospecials}
\def\setupverbatim{\tt
  \def\par{\leavevmode\endgraf} \catcode`\`=\active
  \obeylines \uncatcodespecials \obeyspaces \parindent=5mm \parskip=0pt}
{\obeyspaces\global\let =\ } 
{\catcode`\`=\active \gdef`{\relax\lq}}
\def\beginverbatim{\par\begingroup\setupverbatim\doverbatim}
{\catcode`\|=0 \catcode`\\=12 
  |obeylines|gdef|doverbatim^^M#1\endverbatim{#1|endgroup}}


\newcommand{\luma}{\mbox{$3 \cdot 10^4~$pb$^{-1}$}}
\newcommand{\lumb}{\mbox{$ 10^5~$pb$^{-1}$}}
\newcommand{\lumc}{\mbox{$3 \cdot  10^5~$pb$^{-1}$}}
\newcommand{\tanb}{\mbox{$\tan \beta$}}
\newcommand{\AHbb}{\mbox{$\mathrm{A,H \to b \bar b}$}}
\newcommand{\bbH}{\mbox{$\mathrm{b \bar b H}$}}
\newcommand{\bbAH}{\mbox{$\mathrm{b \bar b A/H}$}}
\newcommand{\mt}{\mbox{$\mathrm{m_t}$}}
\newcommand{\mA}{\mbox{$\mathrm{m_A}$}}
\newcommand{\mH}{\mbox{$\mathrm{m_H}$}}
\newcommand{\mh}{\mbox{$\mathrm{m_h}$}}
\newcommand{\Hhh}{\mbox{$\mathrm{H \to h h }$}}
\newcommand{\hbb}{\mbox{$\mathrm{h \to b \bar b }$}}
\newcommand{\mbb}{\mbox{$\mathrm{m_{b \bar b} }$}}
\newcommand{\mjj}{\mbox{$\mathrm{m_{jj} }$}}
\newcommand{\mbbbb}{\mbox{$\mathrm{m_{b \bar b b \bar b} }$}}
\newcommand{\pT}{\mbox{$\mathrm{p_T}$}}

\begin{titlepage}

\begin{flushright}
{\rm  CERN-TH/97-210}\\
\end{flushright}

\begin{center}

\vspace*{5mm}
\vskip 2mm
 
{\bf \LARGE MSSM Higgs searches in multi-b-jet final states} \\
\vspace{0.5cm}
{\bf \LARGE at the LHC} \\

\end{center}
 
\vspace{5mm}
\begin{center}
  {\bf El\. zbieta Richter-W\c{a}s\footnote{ Supported in part by
 Polish Government grants
 2P03B00212 and 2P03B17210. }}\\
  {\em CERN, TH Division, 1211 Geneva 23, Switzerland}\\
  {\em Institute of Computer Science, Jagellonian University,}\\
  {\em Institute of Nuclear Physics}\\
  {\em 30-059 Krak\'ow, ul.Kawiory 26a, Poland}\\
\vspace{0.5cm}
  {\bf Daniel Froidevaux}\\
  {\em CERN, PPE Division, 1211 Geneva 23, Switzerland}\\
\end{center}

\vspace{0.5cm}
\begin{center}
{\bf Abstract}
\end{center}

  This paper discusses the possibility to observe a signal from MSSM Higgs 
boson decays into final states containing four b-jets. 
Two specific channels are considered:
$\mathrm{b \bar b H}$ and $\mathrm{b \bar b A}$ production with
 $\mathrm{H, A \to b \bar b}$,
for large values of~\mH,~\mA\ and~$\mathrm{\tan \beta}$, and
$\mathrm{H \to hh \to b \bar b b \bar b}$ decays for
 150 GeV~$<$~\mH~$<$~2$\mathrm{m_t}$ and
 for low values of $\mathrm{\tan \beta}$.
Both channels are difficult to extract because of the very large
reducible and irreducible QCD backgrounds.
Even with an ultimate integrated
luminosity, expected per LHC experiment, of \lumc, the  region of 
the MSSM parameter space covered by
these channels does not extend the reach beyond that accessible 
to other channels that were studied in the past. 
Nevertheless, their observation would help in constraining
the couplings and branching ratios of the MSSM Higgs bosons.

\vspace{0.1cm}
\begin{flushleft}
{\bf  CERN-TH/97-210} \\
{\bf  August 1997}
\end{flushleft}
  
\end{titlepage}




\section{Introduction}

One of the most attractive theoretical extensions of the Standard Model (SM)
is the Minimal Supersymmetric Standard Model (MSSM)\cite{MSSM, Guide}.
 Recently, the interest
in this model has grown for several reasons. It is the simplest version of 
low-energy supersymmetry, with the minimal gauge group and particle content
compatible with phenomenology. It preserves all SM predictions at low energies,
and is therefore not contradicted by any of the precision measurements
at LEP \cite{LEP2}. 
It also has a high level of predictivity in the Higgs sector. 
At tree level, all Higgs-boson masses 
and couplings can be expressed in terms of only two parameters, usually chosen
to be the ratio of the vacuum expectation values of the two Higgs 
doublets, $\mathrm{\tan \beta}$, and the mass of the A boson, \mA. 
Radiative corrections \cite{radiative} introduce a dependence of the
 masses and couplings 
on the top-quark mass, the squark masses and the mixing parameters in the 
stop--sbottom sector. All couplings of the MSSM Higgs bosons to fermions 
and bosons can be obtained from the SM couplings by multiplying the latter 
with the appropriate MSSM correction factors, which depend on the 
parameters listed above.

The MSSM predictions concerning the Higgs-boson masses have important
phenomenological consequences. 
One of the Higgs bosons, the scalar h,
is relatively light, with a maximum allowed mass value varying from
105 to 153 GeV, depending on the MSSM parameters and the top-quark mass.
The masses of the other Higgs bosons are expected to be larger and
nearly degenerate for \mA~$>~200$~GeV.

There are a variety of production and decay channels through which 
the MSSM Higgs bosons can be observed at LHC \cite{Note-074}.
Some of these are similar to the SM, such as the $\mathrm{H,~h~\to~\gamma~\gamma}$ 
or~$\mathrm{H \to 4 \ell}$ channels. Others are allowed in the~SM but have large 
enough rates only in the MSSM case, e.g.~$\mathrm{H, A \to \tau \tau,~\mu \mu}$, 
and~$\mathrm{t \bar t}$. There are also channels characteristic of the MSSM, 
such as~$\mathrm{H^{\pm} \to \tau \nu}$,
 $\mathrm{A \to Zh}$, $\mathrm{H \to hh}$, as well as the 
production of h~bosons in squark and gluino decays.

The work described in \cite{Note-074} presents an extended and complete study 
of most of these channels, leading to realistic predictions for the LHC 
discovery potential in the MSSM Higgs sector. 
Figures~\ref{FMSSM:lhcmA} and~\ref{FMSSM:lhcmB} show 
the final results of~\cite{Note-074}, namely the $\mathrm{5\sigma}$-discovery contour
curves, for \mt~=~175~GeV and for integrated luminosities of~\luma\ and~\lumc\ per LHC
 experiment, respectively, which represent the integrated luminosities 
expected after three and ten years of operation.
With the initial, modest integrated luminosity of \luma, a large fraction 
of the parameter space is already covered. With the ultimate integrated 
luminosity of \lumc, the LHC  discovery potential maps
the complete parameter space. For the vast majority of cases, 
the experiment would be able to distinguish between the SM and MSSM cases.

The MSSM Higgs sector is quite challenging experimentally for the  
LHC \cite{bothTP}, since, most often, the signal-to-background ratios 
are much smaller than unity and the detector resolution in several of 
the accessible channels is far from optimal.
This sets stringent requirements on the performance in terms of energy
and momentum resolution and of particle identification; however, on the other hand,
the variety of channels makes this sector an excellent benchmark to 
evaluate the discovery potential of the detector. A very good performance 
of the electromagnetic calorimeter, excellent b-tagging capabilities, 
good $\mathrm{E_T^{miss}}$-resolution and $\mathrm{\tau}$-identification are all crucial 
ingredients to fully explore the MSSM Higgs sector.

Channels with b-jets identified in the final state were considered so far
only if associated with hard leptons or photons: 
$\mathrm{Wh}$ with $\mathrm{h \to b \bar b}$,
 $\mathrm{H/A \to t \bar t \to \ell \nu j j b \bar b}$, 
$\mathrm{A \to Zh \to \ell \ell b \bar b}$ and 
$\mathrm{H \to hh \to b \bar b \gamma \gamma}$.
All of these channels were found to be observable only for 
low values of~$\mathrm{\tan \beta}$. The possibility to improve the sensitivity 
by using identified b-jet spectators was also recently 
considered~\cite{Donatella} in the case of $\mathrm{b \bar bA, b \bar bH}$ production
with $\mathrm{H/A \to \tau \tau}$ decay. In all these cases, however, isolated leptons 
or photons provide a straightforward experimental trigger, and the purely
hadronic background from QCD multi-jet production is not a source for concern.

Final states containing more than two b-jets have not been explored 
so far in these searches. They nevertheless deserve some attention, since 
they correspond to the dominant production and decay modes of the MSSM Higgs
bosons over a large fraction of the parameter space. 
Also, it has been reported in recent theoretical papers~\cite{Gunion} that 
they should be very promising for discovery in the MSSM Higgs sector.

Two possible channels will be discussed in this paper\footnote{
The present paper is  a slightly modified version of 
 \cite{Note-104}.}, 
both leading to final states containing four b-jets but no lepton nor photon for the
experimental trigger.
\begin{itemize}
\item The first channel arises from $\mathrm{b \bar bH}$ and $\mathrm{b \bar bA}$
 production,
which is strongly enhanced for large values of~$\mathrm{\tan \beta}$, since the 
production rates grow like~$\mathrm{\tan^2 \beta}$. These production processes were
studied so far through the $\mathrm{\tau \tau}$ and $\mathrm{\mu \mu}$ decay 
modes over the 
mass range from~$100$ to~$500$~GeV. The $\mathrm{\tau \tau}$ channel requires 
excellent $\mathrm{\tau}$-identification to suppress the huge background of
hadronic jets from various sources, but also excellent $\mathrm{E_T^{miss}}$-resolution
for the reconstruction of the $\mathrm{\tau \tau}$ invariant mass.
The region of parameter space accessible to the  $\mathrm{\tau \tau}$ channel, 
which has a branching ratio of~$\sim$~10\%, is superior to that of the
$\mathrm{\mu \mu}$~channel;  for \mA~=~300~GeV, the
 $\mathrm{5\sigma}$-discovery contour
extends down to~$\mathrm{\tan \beta~\sim~10}$ for an integrated luminosity of \luma, 
and down to~$\mathrm{\tan \beta~\sim~7}$ for ATLAS+CMS combined
 with an integrated luminosity of \lumc\ per experiment. 
For larger masses, the sensitivity degrades rapidly, due to the
decreasing production rates and the degradation of the $\mathrm{\tau \tau}$ mass 
resolution. For \mA~=~500~GeV,
 the accessible region reaches only down 
to~$\mathrm{\tan \beta \sim}$~25 (resp.~$\sim$~18) for an integrated luminosity
of~\luma\ (resp.~\lumc).
The rates expected for the $\mathrm{b \bar b}$~decay mode would be much higher, 
since the branching ratio is about~90\%. However, the observation of
final states containing multiple b-jets is very challenging:
it requires excellent b-tagging performance and very efficient jet 
reconstruction.
\item The second channel is the $\mathrm{H \to hh}$ decay, which is dominant for low
values of~$\mathrm{\tan \beta}$ and for~\mH~$<$~2\mt. 
The observation of this channel would be very interesting, since it would
correspond to the simultaneous discovery of two MSSM Higgs bosons. The final 
state $\mathrm{H \to hh \to bb \gamma \gamma}$ has been studied so far. 
It provides a straightforward experimental trigger and offers good 
kinematical constraints and mass resolution for the reconstruction of~\mH.
The expected rates are, however, very low, even when requiring only one 
identified b-jet in the final state. The sensitivity to this channel decreases 
as~$\mathrm{\tan \beta}$ increases, and the accessible region of parameter space 
extends up to~$\mathrm{\tan \beta~\sim~2.5}$ for an integrated luminosity of~\luma\ and 
up to~$\mathrm{\tan \beta~\sim~4.5}$ for ATLAS+CMS combined with an integrated luminosity
 of \lumc\ per experiment. 
The $\mathrm{H \to hh \to b \bar b b \bar b}$ channel would yield much higher rates, 
by a factor of almost~1000; however, the  4-jet final 
state will be difficult to trigger on and to extract from the huge QCD 
background.
\end{itemize}

The main difficulty and uncertainties in the analysis presented here arise 
from the evaluation of the QCD multi-jet/multi-b-jet backgrounds. 
There are many sources of uncertainties in the predictions for QCD multi-jet 
production: the parton density functions, the choice of scale for the strong 
coupling constant, the calculations of the exact and complete matrix elements
 and the higher-order corrections.
At tree level, matrix elements for all parton processes with at most five 
partons in the final state, have been calculated and embedded in a Monte Carlo
generator called~NJETS~\cite{NJETS}. However, the parton-flavour information 
is not accessible in this generator.
For the studies presented here, the PYTHIA Monte Carlo generator~\cite{PYTHIA},
with its QCD parton-shower modelling and gluon splitting into heavy quarks,
has been used, since the flavour content of the jets in the final state is 
of crucial importance for the evaluation of many of the background processes.
Only leading-order terms are controlled in this approach.
Predictions from both generators for events with at least 2, 3, 4 reconstructed 
high $\mathrm{p_T}$ jets ($\mathrm{p_T^{jet}>}$~30, 40, 50 GeV) were compared
 and results from PYTHIA were found to be systematically lower by a factor 2-3.
The uncertainties on the predictions for high-\pT\ multi-jet final states
are therefore estimated to be as high as at least a factor of~3 and the results presented 
throughout this paper as background rates should be considered as an optimistic.

The expected detector performance is simulated with the ATLFAST 
package~\cite{Note-079}. This package, used extensively 
for fast simulation of the ATLAS detector response provides  reliable  estimates
of the detector response to hadronic jets. Though the  parameterisations of the 
detector performance are ATLAS--specific,
 the results presented here  can be considered as representative of what is to be
 expected at the LHC.
For those results evaluated in  the case of high-luminosity operation,
pile-up effects are included. The reconstructed jet 
energies, after kinematical selection, are recalibrated, 
as explained in~\cite{Note-079}, 
to obtain the  correct peak position for e.g. the \mbb\ distribution. For the 
analysis presented here, a rather optimistic b-tagging performance is assumed
(see~\cite{ID-TDR} for the most recent ATLAS results); for the low-luminosity case,
an overall b-tagging efficiency of~$\mathrm{\epsilon_b~=~60\%}$ is assumed, with a
rejection against c-jets of~$\mathrm{R_c~=~10}$ and against light-quark and gluon jets
of~$\mathrm{R_j~=~100}$. For the high-luminosity case, the b-tagging efficiency is 
degraded  to $\mathrm{\epsilon_b~=~50\%}$, assuming the same rejection 
against non-b-jets.

\begin{Fighere}
\begin{center}
\epsfig{file=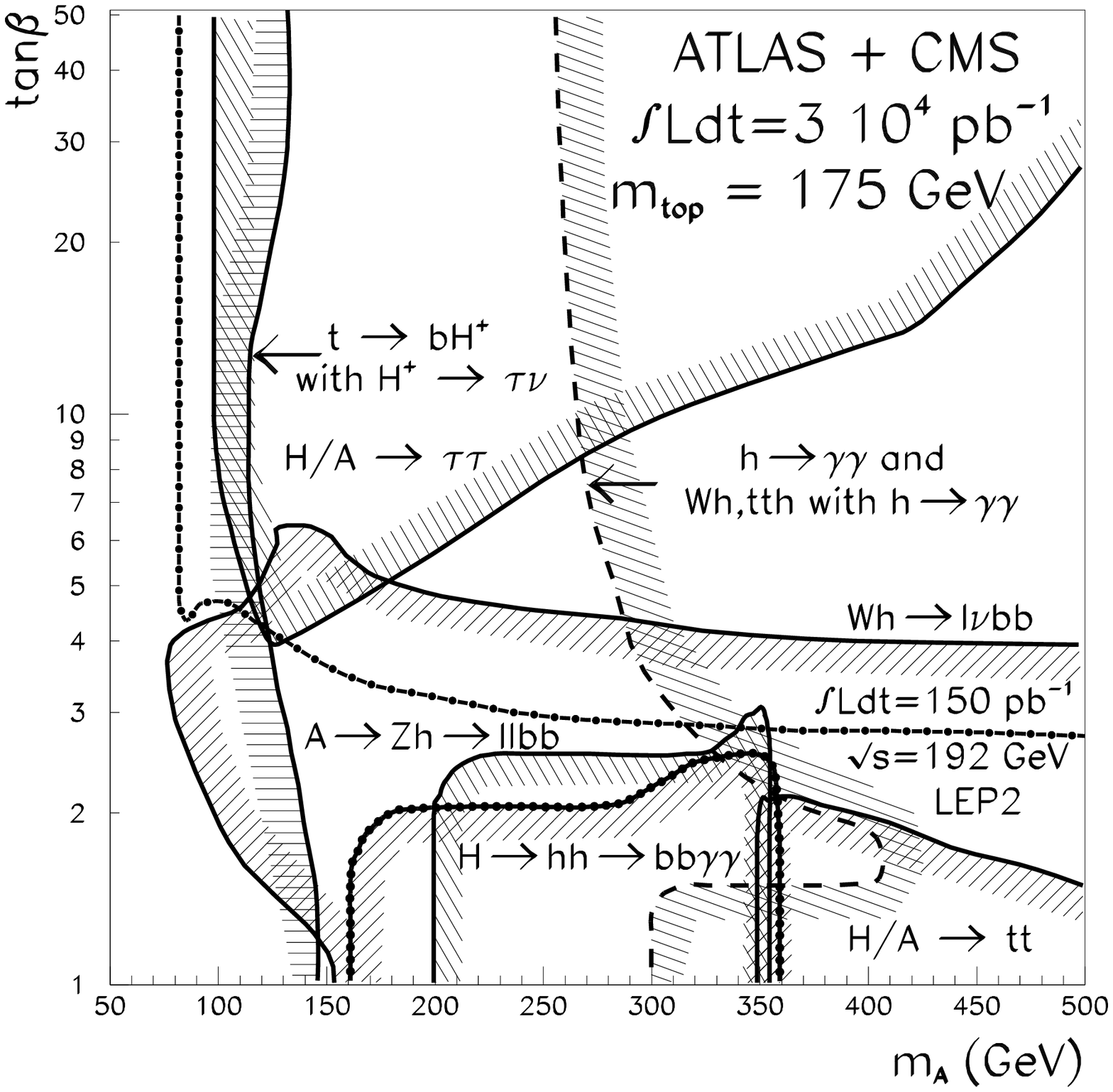,height= 14cm,width=14cm}
\end{center}
\caption{\em 
For \mt~=~175~GeV and an integrated luminosity of~\luma, expected LHC (ATLAS + CMS)
$5\sigma$-discovery contour curves 
in the (\mA,~\tanb) plane for all MSSM Higgs-boson channels studied
in~\cite{Note-074}.
\label{FMSSM:lhcmA}}
\end{Fighere}

\newpage
\begin{Fighere}
\begin{center}
\epsfig{file=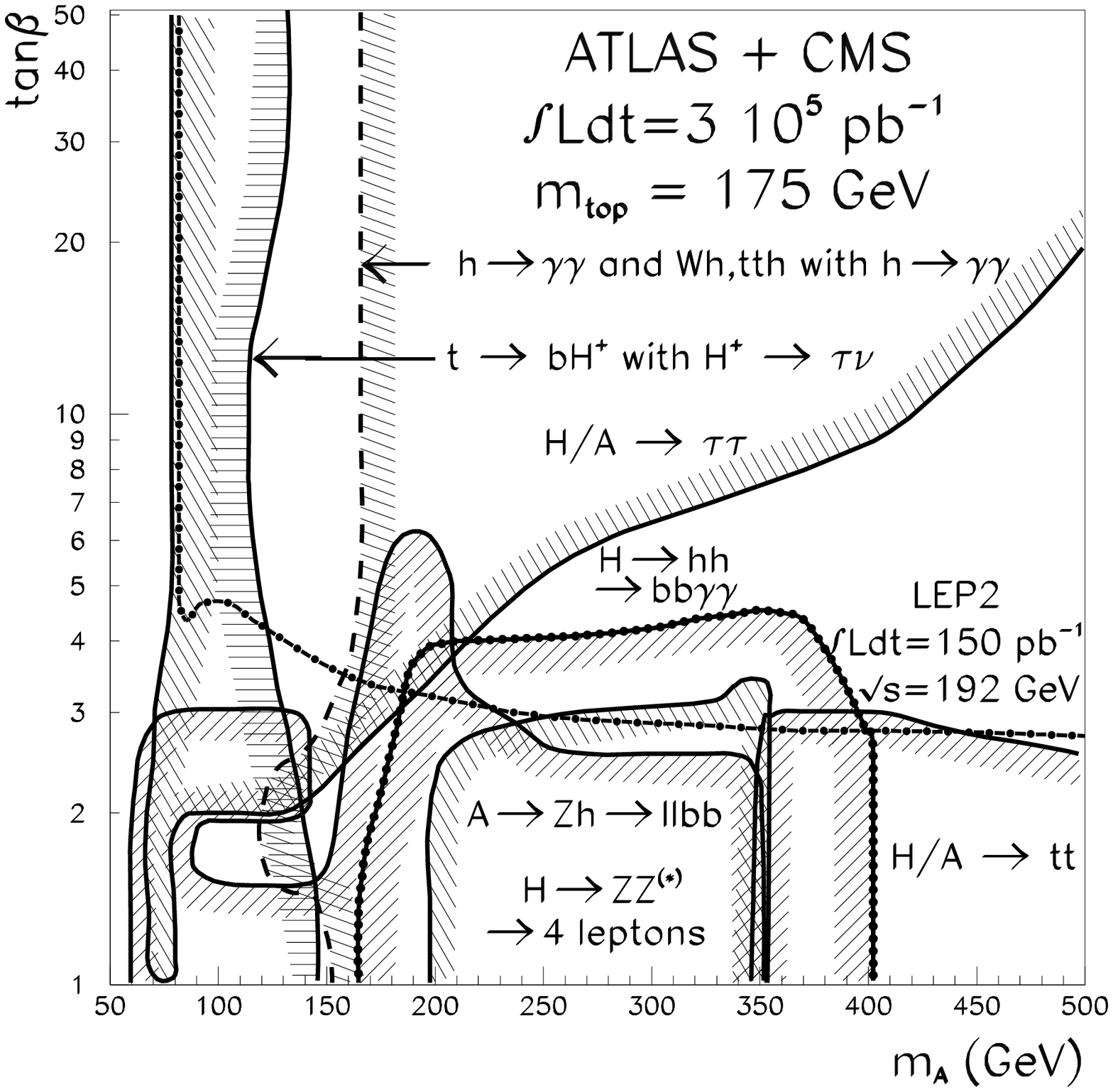,height=14cm,width=14cm}
\end{center}
\caption{\em 
For \mt~=~175~GeV and an integrated luminosity of~\lumc, expected LHC (ATLAS + CMS)
$5\sigma$-discovery contour curves 
in the (\mA,~\tanb) plane for all MSSM Higgs-boson channels studied
in~\cite{Note-074}.
\label{FMSSM:lhcmB}}
\end{Fighere}

\newpage
Two questions are of particular interest in the context of  these studies:

$\bullet$ Is it possible, using the $\mathrm{b \bar b H/A}$ channel with
 $\mathrm{H/A \to b \bar b}$, 
to extend the region in parameter space already  covered by the 
$\mathrm{\tau \tau}$ channel, particularly to values of~\mA\ larger than~500~GeV?

$\bullet$ Is it possible, using the $\mathrm{H \to hh \to b \bar b b \bar b}$ channel,
to extend the region in parameter space already covered by 
the $\mathrm{H \to hh \to b \bar b \gamma \gamma}$ channel?

 \boldmath
\section{Observability of $\mathrm{H/A \to b \bar b}$ for large $\mathrm{\tan \beta}$}
\unboldmath

 \subsection{Signal events}

For large values of~\mH\ and~\mA, the final state in this channel has a 
very characteristic topology: the two hardest jets in the event come from 
the $\mathrm{H/A \to b \bar b}$ decay, while the softer ones come from the 
associated $\mathrm{b \bar b}$ production and from initial/final-state radiation. 
These features have been used for the event selection. At least four 
reconstructed jets are required in the final state, with the three or four 
jets of highest transverse energy tagged as b-jets. The two jets with the
largest transverse energy are used for the reconstruction of the
Higgs-boson mass,~\mbb. The selection criteria have been optimised for each 
value of~\mA\ and~\mH, chosen from~300 to~900~GeV, as follows:
\begin{itemize}
\item at least four reconstructed jets (ordered in decreasing~\pT) 
with~$\mathrm{\pT^{j_3}~>~50}$~GeV and~$\mathrm{\pT^{j_4}~>}$~30~GeV;
 the thresholds on the 
two hardest jets are raised to:

$\bullet$ $\mathrm{\pT^{j_1}~>}$~100~GeV and $\mathrm{\pT^{j_2}~>}$~70~GeV 
          (selection $\mathrm{S_{300}}$);

$\bullet$ $\mathrm{\pT^{j_1}~>}$~200~GeV and
 $\mathrm{\pT^{j_2}~>}$~100~GeV (selection $\mathrm{S_{500}}$);

$\bullet$ $\mathrm{\pT^{j_1}~>}$~250~GeV and
 $\mathrm{\pT^{j_2}~>}$~150~GeV (selection $\mathrm{S_{700}}$);

$\bullet$ $\mathrm{\pT^{j_1}~>}$~300~GeV and
 $\mathrm{\pT^{j_2}~>}$~200~GeV (selection $\mathrm{S_{900}}$);

\item  the three highest-\pT\ jets are required to be tagged as b-jets; an
additional tagged b-jet is required in two different approaches:

$\bullet$ in algorithm A, the fourth highest-\pT\ jet is chosen;

$\bullet$ in algorithm B, any tagged b-jet among the additional 
jets is accepted;

\item the two highest-\pT\ jets are used to reconstruct 
the \mbb\ distribution;

\item the events are accepted if \mbb\ falls within~$\mathrm{\pm}$80~to~100~GeV
of the considered Higgs-boson mass; the mass window was chosen to accept 
approximately~70\% of the signal events.
\end{itemize}

\newpage
The acceptance of the selection $S_{500}$ increases from~10\% to~40\% 
as the Higgs-boson mass increases from~500 to~900~GeV, as shown
in~Table~\ref{THIGGS:efficiency1}. The three highest-\pT\ jets are found
to be true b-jets only in~6\% to~14\% of the events, due to initial/final-state
radiation. This proportion decreases to~1.6\% to~3.2\% of the events if the 
fourth highest-\pT\ jet is also required to be a true b-jet (algorithm~A). 
For algorithm~B, the acceptance is improved by about~60\%. 
These rather low acceptances for the signal events are unavoidable
to reduce the huge QCD backgrounds to an acceptable level, as discussed 
in~Section~2.2. It was checked for example that the fact of  requiring any 
combination of the reconstructed jets to be identified as b-jets leads
to much worse signal-to-background ratios and significances.

Table~\ref{THIGGS:efficiency2} shows, for an integrated luminosity of~\lumc, 
the expected rates of signal events after applying the selection~$S_{500}$ 
and the b-tagging procedure, for different assumptions on the b-tagging
performance in addition to the default one described above. The signal sample 
is dominated by events containing true b-jets: the contribution from events 
with at least one mis-identified jet is below~10\%, as shown by comparing 
the results for the default jet rejections with those for an almost infinite
rejection,~R~=~10$^6$. 

\begin{Fighere}
\vspace{-1cm}
\begin{center}
\epsfig{file=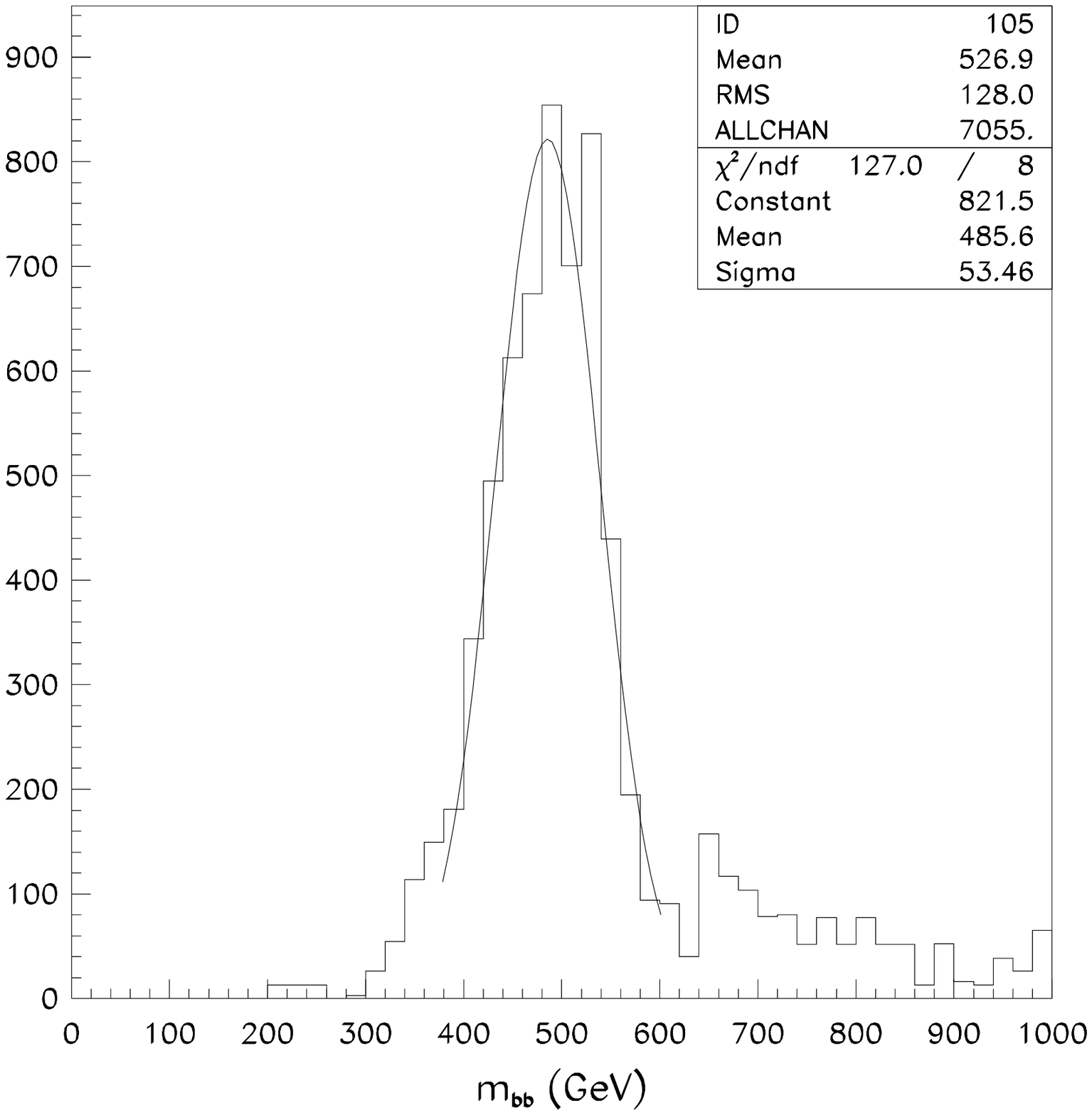,height=10cm,width=10cm}
\end{center}
\caption{\em 
Reconstructed $\mbb$ mass after applying the selection $S_{500}$, for 
the $b \bar b H$ signal with $H \to b \bar b$ decay and for~\mH~=~500~GeV.
The degradation in resolution expected from pile-up at high luminosity has been
included.
\label{FHIGGS:hbb}}
\vspace{2mm}
\end{Fighere}

                                  \begin{Tabhere}
      \newcommand{\lstrut}{{$\strut\atop\strut$}}
             \caption {\em Acceptance of selection $S_{500}$ and fractions of
                           events containing one or more true b-jets, for three
                           values of \mA\ and~\mH.
                           \label{THIGGS:efficiency1}}
                                     \vspace{2mm}
                                   \begin{center}
       \begin{tabular}{|c||c||c|c|c|c||c|c|}
                                    \hline 
 $\mathrm{b \bar bH, b \bar bA}$  &$\mathrm{\sigma}$ & Four & Jet1 & + jet2 & + jet3 & 
                                                      Algorithm & Algorithm  \\
 with $\mathrm{H/A \to b \bar b}$  & (pb)    & jets & =~true  & =~true     & =~true     & 
                                                          A     &     B      \\ 
 Selection $\mathrm{S_{500}}$    &         &      &  b-jet & b-jet    & b-jet    &  
                                                                &            \\ 
                                    \hline \hline
 \mH~=~500~GeV & 3.5  & 10.8\% & 10.0\% & 8.4\% & 5.6\% & 1.6\% & 2.4 \% \\
 \mH~=~700~GeV & 0.9  & 20.4\% & 18.0\% & 14.8\% & 8.0\% & 2.0\% & 3.2 \%\\
 \mH~=~900~GeV & 0.2  & 39.2\% & 36.0\% & 26.4\% & 13.6\% & 3.2\% & 5.5 \% \\
\hline 
\end{tabular}                                                                   
\end{center}                                                                    
                                    \end{Tabhere}

                                  \begin{Tabhere}
      \newcommand{\lstrut}{{$\strut\atop\strut$}}
             \caption {\em Expected rates of reconstructed signal events
                           after applying selection~$\mathrm{S_{500}}$ and b-tagging,
                           for $\mathrm{\tan \beta~=~30}$ and for an integrated 
                           luminosity of \lumc\ (ATLAS). 
The results are shown for three different assumptions on the b-tagging performance.
                           \label{THIGGS:efficiency2}}
                                     \vspace{2mm}
                                   \begin{center}
       \begin{tabular}{|c||c||c|c|c||c|c|}
                                    \hline 
 Selection $\mathrm{S_{500}}$ & Four & Jet1     & + jet2   & + jet3   & 
                                                     Algorithm & Algorithm  \\
                     & jets & tagged   & tagged   & tagged   & 
                                                         A     &     B      \\ 
 \mA,~\mH            &      & as b-jet & as b-jet & as b-jet &   
                                                               &            \\ 
                                    \hline \hline
\cline{3-7} \multicolumn{2}{|c||}{}& \multicolumn{5}{|c|}{ $\mathrm{\epsilon_b~=~0.5}$,
 \ \ 
$\mathrm{\epsilon_c~=~10^{-6}}$, \ \ R~=~$10^6$ } \\
                                    \hline \hline
 500 GeV & 
  111000   &  52400  &  22000  & 7250  &  1090    &   1670  \\
 700 GeV &
   54600   &  24700  &   9850  & 2660  &   363    &    572  \\
 900 GeV &
   23800   &  10700  &   4030  & 1030  &   125    &    217  \\
\hline 
\cline{3-7} \multicolumn{2}{|c||}{}& \multicolumn{5}{|c|}{ $\mathrm{\epsilon_b~=~0.5}$, \ \ 
$\mathrm{ \epsilon_c~=~0.1}$, \ \ R~=~$10^2$ } \\
                                    \hline \hline
 500 GeV & 
  111000   &  52500  &  22100  & 7400  &  1180    &   1770  \\
 700 GeV &
   54600   &  24700  &   9940  & 2750  &   404    &    622  \\
 900 GeV &
   23800   &  10700  &   4080  & 1070  &   142    &    238  \\
\hline 
\cline{3-7} \multicolumn{2}{|c||}{}& \multicolumn{5}{|c|}{ $\mathrm{\epsilon_b~=~0.6}$, \ \ 
$ \mathrm{\epsilon_c~=~0.1}$, \ \ R~=~$10^2$ } \\
                                    \hline \hline
 500 GeV & 
  111000   &  63000  &  31900  & 12700 &  2420    &   3630  \\
 700 GeV &
   54600   &  29700  &  14300  & 4730  &   824    &   1270  \\
 900 GeV &
   23800   &  12900  &   5860  & 1830  &   288    &    486  \\
\hline 
\end{tabular}                                                                   
\end{center}                                                                    
                                    \end{Tabhere}

The reconstructed  \mbb\ peak is rather broad, because final-state radiation
and hadronisation degrade the mass resolution and also because a significant
fraction (about~20\%) of the $\mathrm{b \bar b}$~combinations entering the distribution 
are incorrect (one or both b-jets do not originate from the 
$\mathrm{H \to b \bar b}$ decay). Figure~\ref{FHIGGS:hbb} shows as an example the
distribution of the reconstructed \mbb\ for \mH~=~500~GeV; the acceptance 
in the chosen mass bin (\mH~$\pm$~80~GeV) is about~70\%.

The quality of the mass reconstruction has been studied separately, 
using the $\mathrm{gg \to H \to b \bar b}$ process, which does not suffer from the
combinatorial backgrounds of the $\mathrm{b \bar b H}$ process. The mass resolution obtained
and the acceptance in the mass window are shown in~Table~\ref{THIGGS:resolution}
and in~Fig.~\ref{FHIGGS:hbb2} for all events without any selection.
The large tails in the mass distributions that can be seen 
in~Fig.~\ref{FHIGGS:hbb2} are almost completely removed if one applies
the selection criteria on the jet transverse energies, but the events in the 
tails are nevertheless lost. Most of these losses are accounted for in the
acceptances quoted in Table~\ref{THIGGS:efficiency1}. Finally,
Table~\ref{THIGGS:signaln} shows the total expected signal rates for three 
values of~\mH\ and the appropriate selection cuts, before and after applying 
the chosen \mbb~mass window (including pile-up at high luminosity), 
for an integrated luminosity of~\lumc.

                                  \begin{Tabhere}
      \newcommand{\lstrut}{{$\strut\atop\strut$}}
             \caption {\em Mass resolutions expected for the reconstructed 
                       \mbb\ mass, for the $\mathrm{gg \to H \to b \bar b}$ process 
                       in the absence and presence of pile-up. No selection
                       cuts are applied.
                       \label{THIGGS:resolution}}
                                     \vspace{2mm}
                                   \begin{center}
       \begin{tabular}{|c||c|c||c|c|}
                                    \hline 
\cline{2-5} \multicolumn{1}{|c||}{ Higgs mass (GeV)}& \multicolumn{2}{|c||}{No pile-up } &
\multicolumn{2}{|c|}{With  pile-up } \\
                                    \hline
 & $\sigma_m$ (GeV) & Within $\mathrm{\pm~2~\sigma_m}$ & $\mathrm{\sigma_m}$ (GeV)
 & Within $\mathrm{\pm~2~\sigma_m}$ \\
\hline \hline
  300     & 22.7  & 62\%    & 26.4   &  60\%   \\
  500     & 33.8  & 57\%    & 39.7   &  55\%  \\
  700     & 38.3  & 52\%    & 44.0   &  50\%   \\   
  900     & 52.6  & 53\%    & 63.5   &  52\%   \\ 
\hline  
\end{tabular}                                                                   
\end{center}                                                                    
                                    \end{Tabhere}

                                  \begin{Tabhere}
      \newcommand{\lstrut}{{$\strut\atop\strut$}}
             \caption {\em Expected numbers of signal events after applying
                       the appropriate selection criteria, for an integrated
                       luminosity of \lumc\ (ATLAS). The numbers of events accepted 
                       within the \mbb~mass window are shown in brackets.
                       \label{THIGGS:signaln}}
                                     \vspace{2mm}
                                   \begin{center}
       \begin{tabular}{|c|c||c|c|c|}
                                    \hline 
  $\mathrm{b \bar b H, b \bar b A}$ production             &Kinematical & Three  & Algorithm & Algorithm   \\
 with $\mathrm{H/A \to b \bar b}$              &   cuts     & b-jets &     A     &     B       \\
                                    \hline \hline
 \mH~=~500~GeV       & $1.1 \cdot 10^5$  &  7400   & 1100   &  1800   \\
 Selection $\mathrm{S_{500}}$ &                   &  (5000)  & (960)  & (1300)  \\ \hline
 \mH~=~700~GeV       & $3.4 \cdot 10^4$  &  2289   &  360   &   530   \\
 Selection $\mathrm{S_{700}}$ &                   & (1530)  & (240)  &  (350)  \\ \hline
 \mH~=~900~GeV       & $1.3 \cdot 10^4$  &   900   &  135   &   210   \\
 Selection $\mathrm{S_{900}}$ &                   &  (524)  &  (80)  &  (122)  \\ \hline
\end{tabular}                                                                   
\end{center}                                                                    
                                    \end{Tabhere}

\begin{Fighere}
\vspace{-1cm}
\begin{center}
\epsfig{file=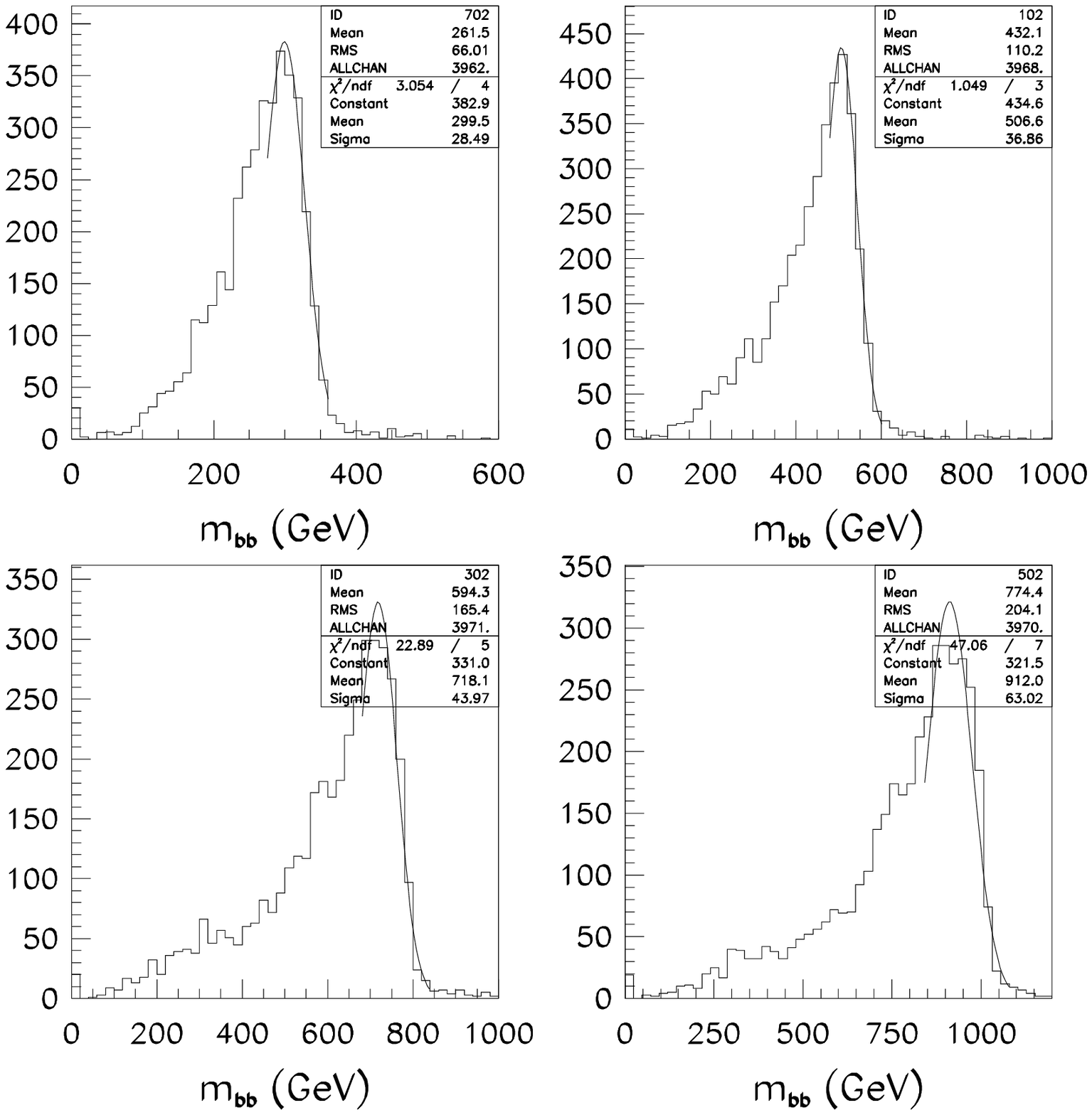,height=10cm,width=10cm}
\end{center}
\caption{\em 
Reconstructed \mbb\  before any selection cuts, for 
the $\mathrm{gg \to  H \to b \bar b}$ process and for \mH~=~300,~500,~700 and~900~GeV. 
The degradation in resolution expected from pile-up at high luminosity has
been included.
\label{FHIGGS:hbb2}}
\end{Fighere}

 \subsection{Background events}

The background to the  $\mathrm{b \bar b H/A}$ signal with $\mathrm{H/A \to b \bar b}$ decay is
completely dominated by QCD multi-jet processes with final states
containing a variable number of real b-jets. For the analysis presented here,
large samples of background events were generated using PYTHIA, based
on the hard-scattering di-jet sub-processes. The hard-scattering process
was accompanied by initial-state radiation, which provides one of the
sources of additional jets; gluon splitting and/or final-state radiation
also provide a source of additional jets, some of which originate from b-quarks.
   
The event generation was organised in several \pT~bins with different 
statistics, typically $5 \cdot 10^4$ events to $4 \cdot 10^6$ events per bin.
A sufficiently large number of events were generated in each bin to obtain
statistically significant samples after applying the selection criteria for
the final states of interest. 

As for the signal, the background events were simulated through the 
ATLFAST~package, with pile-up effects included for the high-luminosity case.
The b-tagging procedure and the jet energy recalibration were also applied.
Table~\ref{TBDG:efficiency1} shows the summed production cross-sections and the 
kinematical acceptances for selection $S_{500}$ (before applying the b-tagging
procedure), for the most prominent background sub-processes: $\mathrm{b \bar b}$, 
$\mathrm{q \bar q}$, gb, gq and gg. The kinematical acceptances are about~0.13\%
for all the di-jet sub-processes, but the initial cross-sections vary over 
several orders of magnitude. Figure~\ref{FHIGGS:backga} shows the contributions
from the various \pT\ bins chosen at generation to the distribution of~\mbb. 
The relative contributions from the different bins vary rapidly depending on 
the selection criteria, so the use of this algorithm for  event generation 
turned out to be rather efficient in this case\footnote{Many thanks to~F. Paige
for this suggestion.}. As an example, even though the production 
cross-sections in the first two bins are different by several orders of 
magnitude ($2~\cdot~10^7$~pb and~$3.5~\cdot~10^4$~pb respectively), the
contributions to the total background from these bins after selection $\mathrm{S_{500}}$
are comparable.

After selection, but before applying the b-tagging procedure, the inclusive 
background rates are approximately a factor of~$10^4$ to~$10^5$ higher than 
the signal rates in the mass bin of interest. Requiring at least three 
identified b-jets reduces this factor to~$10^2$ to~$10^3$. Requiring 
in addition a fourth identified b-jet gains another factor of about~2.
In the sample containing at least three identified b-jets, approximately~40\% 
of the background events contain at least three true b-jets, whereas in the
sample containing at least four identified b-jets this fraction increases
to~65\%. These fractions can be estimated directly from the numbers shown
in~Table~\ref{TBDG:efficiency2} by comparing the expected rates for an almost
infinite rejection against non-b-jets (R~=~$10^6$) with the default
value of this rejection (R~=~$10^2$). 

The dominant remaining background arises 
from~gb and~gg production with gluon splitting into a $\mathrm{b \bar b}$~pair. 
The contribution from direct $\mathrm{gg \to b \bar b}$ production is found to be only
at the level of~$\sim$~10\% of the total background.  
The contribution from $\mathrm{t \bar t}$~production is very small~($\sim$~1\%). 
Figure~\ref{FHIGGS:backgb} shows the respective contributions from the remaining
background processes to the reconstructed \mbb\~mass distribution, namely
from direct~$\mathrm{b \bar b}$, from the summed $\mathrm{b \bar b}$~+~g b and
from the summed all sub-processes, after applying the 
b-tagging procedure with the performance expected at high luminosity.

                                  \begin{Tabhere}
      \newcommand{\lstrut}{{$\strut\atop\strut$}}
             \caption {\em Production cross-sections and kinematical 
                        acceptances of the selection cuts described 
                        in~Section~2.1, for various QCD background processes.
                       \label{TBDG:efficiency1}}
                                     \vspace{2mm}
                                   \begin{center}
       \begin{tabular}{|c||c||c|}
                                    \hline 
 Sub-process & $\mathrm{\sigma}$ (pb)        &  Acceptance for     \\
             & $\mathrm{p_T^{gen}~>~50}$~GeV & selection $\mathrm{S_{500}}$ \\
                  \hline \hline
$\mathrm{gg,qq \to gg,qq}$           & $ 1.32 \cdot 10^7$ & 0.12\%    \\
$\mathrm{gq \to gq}$                 & $ 8.23 \cdot 10^6$ & 0.16\%    \\
$\mathrm{gb \to gb}$                 & $ 4.40 \cdot 10^5$ & 0.15\%    \\
$\mathrm{gg, q \bar q \to q \bar q}$ & $4.32 \cdot 10^5$  & 0.15 \%   \\ 
$\mathrm{gg, q \bar q \to b \bar b}$ & $ 8.46 \cdot 10^4$ & 0.14\%    \\
$\mathrm{t \bar t}$                & $5.05 \cdot 10^2$  & 4.20\%    \\ \hline \hline
Total                       &  $2.22 \cdot 10^7$ & 0.13\%    \\
\hline 
\end{tabular}                                                                   
\end{center}                                                                    
                                    \end{Tabhere}

\begin{Fighere}
\begin{center}
\epsfig{file=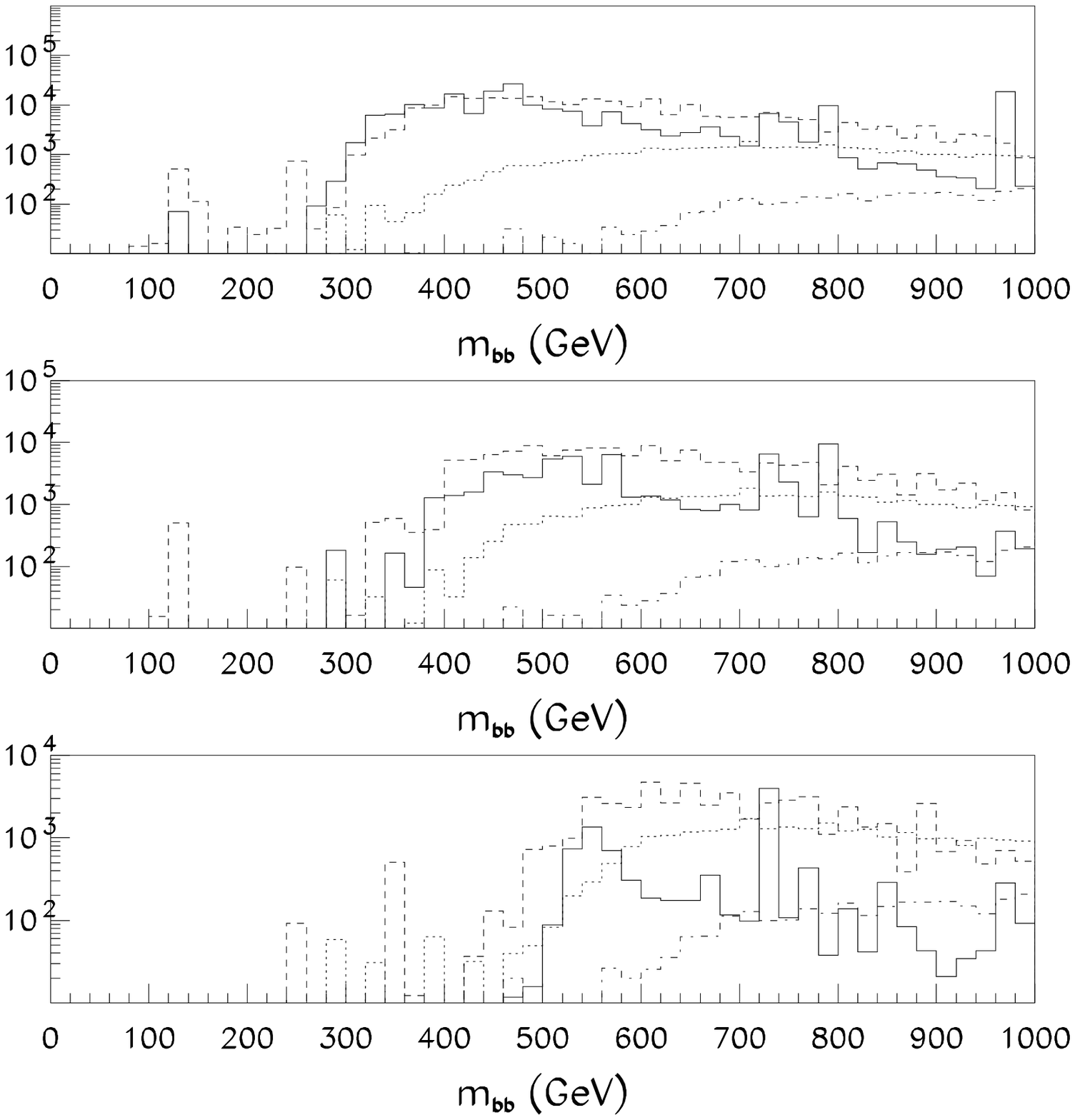,height=10cm,width=10cm}
\end{center}
\caption{\em 
Distribution of the reconstructed \mbb~mass for background events after 
applying the selection cuts~$\mathrm{S_{500}}$~(top),
$\mathrm{S_{700}}$~(middle) and~$\mathrm{S_{900}}$~(bottom), for an integrated luminosity 
of~\lumc\ (ATLAS). The contributions from the different \pT~bins used at generation
are shown separately: $\mathrm{50~<~\pT^{gen}~<~200}$~GeV~(solid), 
$\mathrm{200~<~\pT^{gen}~<~350}$~GeV~(dashed), $\mathrm{350~<~\pT^{gen}~<~500}$~GeV~(dots),
and $\mathrm{\pT^{gen}~>~500}$~GeV~(dot-dashed).
\label{FHIGGS:backga}}
\end{Fighere}

\vspace{0.5cm}

The background estimates presented above explain why the theoretical
estimates of~\cite{Gunion} are much too optimistic. These estimates were
based on an evaluation of the QCD~background in final states containing three
b-jets using only the contribution from the direct $\mathrm{gg \to b \bar b}$ 
sub-process. As appears clearly in~Table~\ref{TBDG:efficiency2},
this sub-process contributes not more than 10~to~15\% of the total 
background for final states containing at least three identified b-jets.
This conclusion was found to be rather independent of the selection cuts
applied to the reconstructed jets, as demonstrated by the results 
of~Section~3.1 in~Table~\ref{TBGD:efficiency5}.

Finally,
Table~\ref{THIGGS:backgroundn} shows the total expected background for the 
different selection cuts, before and after applying the chosen \mbb~window, 
for an integrated luminosity of~\lumc\ and the ATLAS experiment.

\newpage
                                  \begin{Tabhere}
      \newcommand{\lstrut}{{$\strut\atop\strut$}}
      \vspace{-8mm}
             \caption {\em Expected rates of reconstructed background events
                           after selection $\mathrm{S_{500}}$ as a function of the 
                           b-tagging performance,
                           for an integrated luminosity of~\lumc\ (ATLAS).
                       \label{TBDG:efficiency2}}
                                   \begin{center}
       \begin{tabular}{|c||c|c|c|c||c|c|}
                                    \hline 
Sub-process         & Four & Jet1     & + jet2   & + jet3   &  
                                                    Algorithm & Algorithm \\
                    & jets & tagged   & tagged   & tagged   & 
                                                        A     &     B     \\ 
Selection $S_{500}$ &      & as b-jet & as b-jet & as b-jet &
                                                              &           \\ 
                                    \hline \hline
\cline{3-7} \multicolumn{2}{|c||}{}& \multicolumn{5}{|c|}{ $\mathrm{\epsilon_b~=~0.5}$, 
\ \ $\mathrm{\epsilon_c~=~10^{-6}}$, \ \ R~=~$10^6$ } \\
                                    \hline \hline
$\mathrm{gg, q \bar q \to b \bar b}$   &
$3.5 \cdot 10^7$ & $1.4 \cdot 10^7$ & $3.7 \cdot 10^6$ & $7.2 \cdot 10^4$ & $7.3 \cdot 10^3$ & $1.5 \cdot 10^4$ \\
$\mathrm{gg, q \bar q \to q \bar q}$   &
$2.0 \cdot 10^8$ & $1.6 \cdot 10^7$ & $5.2 \cdot 10^6$ & $3.7 \cdot 10^4$ & $3.9 \cdot 10^3$ & $1.0 \cdot 10^4$ \\
$\mathrm{gb \to gb}$                   &
$1.8 \cdot 10^8$ & $4.5 \cdot 10^7$ & $4.2 \cdot 10^6$ & $2.3 \cdot 10^5$ & $2.2 \cdot 10^4$ & $4.1 \cdot 10^4$ \\
$\mathrm{gq \to gq}$                   &
$3.9 \cdot 10^9$ & $3.5 \cdot 10^7$ & $5.5 \cdot 10^5$ & $2.2 \cdot 10^4$ & $1.0 \cdot 10^2$ & $8.4 \cdot 10^2$ \\
$\mathrm{gg,qq \to gg,qq}$             & 
$4.8 \cdot 10^9$ & $8.2 \cdot 10^7$ & $4.6 \cdot 10^6$ & $1.8 \cdot 10^5$ & $1.4 \cdot 10^4$ & $5.2 \cdot 10^4$ \\
 \hline
Total                         &  
$9.1 \cdot 10^9$ & $1.9 \cdot 10^8$ & $1.8 \cdot 10^7$ & $5.4 \cdot 10^5$ & $4.7 \cdot 10^4$ & $1.2 \cdot 10^5$ \\
\hline
\cline{3-7} \multicolumn{2}{|c||}{}& \multicolumn{5}{|c|}{ $\mathrm{\epsilon_b~=~0.5}$,
 \ \ $\mathrm{\epsilon_c~=~0.1}$, \ \ R~=~$10^2$ } \\
                                    \hline \hline
$\mathrm{gg, q \bar q \to b \bar b}$   &
$3.5 \cdot 10^7$ & $1.4 \cdot 10^7$  & $3.9 \cdot 10^6$  & $1.5 \cdot 10^5$ & $1.0 \cdot 10^4$ & $1.9 \cdot 10^4$ \\
$\mathrm{gg, q \bar q \to q \bar q}$   &   
$2.0 \cdot 10^8$ & $2.0 \cdot 10^7$  & $5.4 \cdot 10^6$  & $1.2 \cdot 10^5$ & $6.1 \cdot 10^3$ & $1.5 \cdot 10^4$ \\
$\mathrm{gb \to gb}$                   & 
$1.8 \cdot 10^8$ & $4.6 \cdot 10^7$  & $5.1 \cdot 10^6$  & $3.9 \cdot 10^5$ & $3.1 \cdot 10^4$ & $5.5 \cdot 10^4$ \\
$\mathrm{gq \to gq}$                   &
$3.9 \cdot 10^9$ & $9.2 \cdot 10^7$  & $2.7 \cdot 10^6$  & $1.3 \cdot 10^5$ & $7.0 \cdot 10^3$ & $1.3 \cdot 10^4$\\
$\mathrm{gg,qq \to gg,qq}$             &
$4.8 \cdot 10^9$ & $1.5 \cdot 10^8$  & $7.8 \cdot 10^6$  & $5.0 \cdot 10^5$ & $2.8 \cdot 10^4$ & $7.8 \cdot 10^4$ \\
 \hline
Total                         &
$9.1 \cdot 10^9$  & $3.2 \cdot 10^8$  & $2.5 \cdot 10^7$  & $1.3 \cdot 10^6$ & $8.2 \cdot 10^4$ & $1.8 \cdot 10^5$ \\ 
\hline 
\cline{3-7} \multicolumn{2}{|c||}{}& \multicolumn{5}{|c|}{ $\epsilon_b~=~0.6$,
 \ \ $\epsilon_c~=~0.1$, \ \ R~=~$10^2$ } \\
                                    \hline \hline
$\mathrm{gg, q \bar q \to b \bar b}$   &
$3.5 \cdot 10^7$  & $1.7 \cdot 10^7$  & $5.5 \cdot 10^6$  & $2.3 \cdot 10^5$  & $2.0 \cdot 10^4$ & $3.7 \cdot 10^4$ \\
$\mathrm{gg, q \bar q \to q \bar q}$   &   
$2.0 \cdot 10^8$  & $2.3 \cdot 10^7$  & $7.7 \cdot 10^6$  & $1.8 \cdot 10^5$  & $1.1 \cdot 10^4$  & $2.8 \cdot 10^4$ \\
$\mathrm{gb \to gb}$                   &
$1.8 \cdot 10^8$  & $5.5 \cdot 10^7$  & $7.1 \cdot 10^6$ & $6.3 \cdot 10^5$  & $6.1 \cdot 10^4$  & $1.1 \cdot 10^5$ \\
$\mathrm{gq \to gq}$                   &
$3.9 \cdot 10^9$  & $9.9 \cdot 10^7$  & $3.2 \cdot 10^6$  & $1.9 \cdot 10^5$  & $1.1 \cdot 10^4$  & $2.0 \cdot 10^4$ \\
$\mathrm{gg,qq \to gg,qq}$             &
$4.8 \cdot 10^9$  & $1.7 \cdot 10^8$  & $1.0 \cdot 10^7$  & $7.5  \cdot 10^5$ & $5.0 \cdot 10^4$  & $1.5 \cdot 10^5$ \\ \hline
 Total                        &
$9.1 \cdot 10^9$ & $3.6 \cdot 10^8$  & $3.4 \cdot 10^7$  & $2.0 \cdot 10^6$  & $1.5 \cdot 10^5$  & $3.4 \cdot 10^5$ \\
\hline 
\end{tabular}                                                                   
\end{center}                                                                    
                                    \end{Tabhere}

                                  \begin{Tabhere}
      \newcommand{\lstrut}{{$\strut\atop\strut$}}
      \vspace{-7mm}
             \caption {\em Expected rates of reconstructed background events
                       after applying
                       the appropriate selection criteria, for an integrated
                       luminosity of \lumc\ (ATLAS). The numbers of events accepted 
                       within the chosen \mbb~mass window are shown in brackets.
                       \label{THIGGS:backgroundn}}
                                   \begin{center}
       \begin{tabular}{|c|c||c|c|c|}
                                    \hline 
               & Kinematical  & Three  & Algorithm & Algorithm   \\
               &    cuts      & b-jets &     A     &     B       \\
                                    \hline \hline
 Selection $\mathrm{S_{500}}$ & 
$9.2 \cdot 10^9$ & $1.3 \cdot 10^6$   & $8.2 \cdot 10^4$   & $1.8 \cdot 10^5$ \\ 
             &   & ($3.9 \cdot 10^5$) & ($3.3 \cdot 10^4$) &($6.0 \cdot 10^4$)\\
\hline
 Selection $\mathrm{S_{700}}$ & 
$5.1 \cdot 10^9$ & $7.2 \cdot 10^5$   & $4.0 \cdot 10^4$   & $1.2 \cdot 10^5$ \\ 
             &   & ($1.7 \cdot 10^5$) & ($1.0 \cdot 10^4$) &($2.7 \cdot 10^4$)\\
\hline
 Selection $\mathrm{S_{900}}$ & 
$2.0 \cdot 10^9$ & $2.9 \cdot 10^5$   & $1.8 \cdot 10^4$   & $3.0 \cdot 10^4$ \\
             &   & ($4.8 \cdot 10^4$) & ($2.8 \cdot 10^3$) &($5.2 \cdot 10^3$)\\
\hline
\end{tabular}                                                                   
\end{center}                                                                    
                                    \end{Tabhere}

\begin{Fighere}
\begin{center}
\epsfig{file=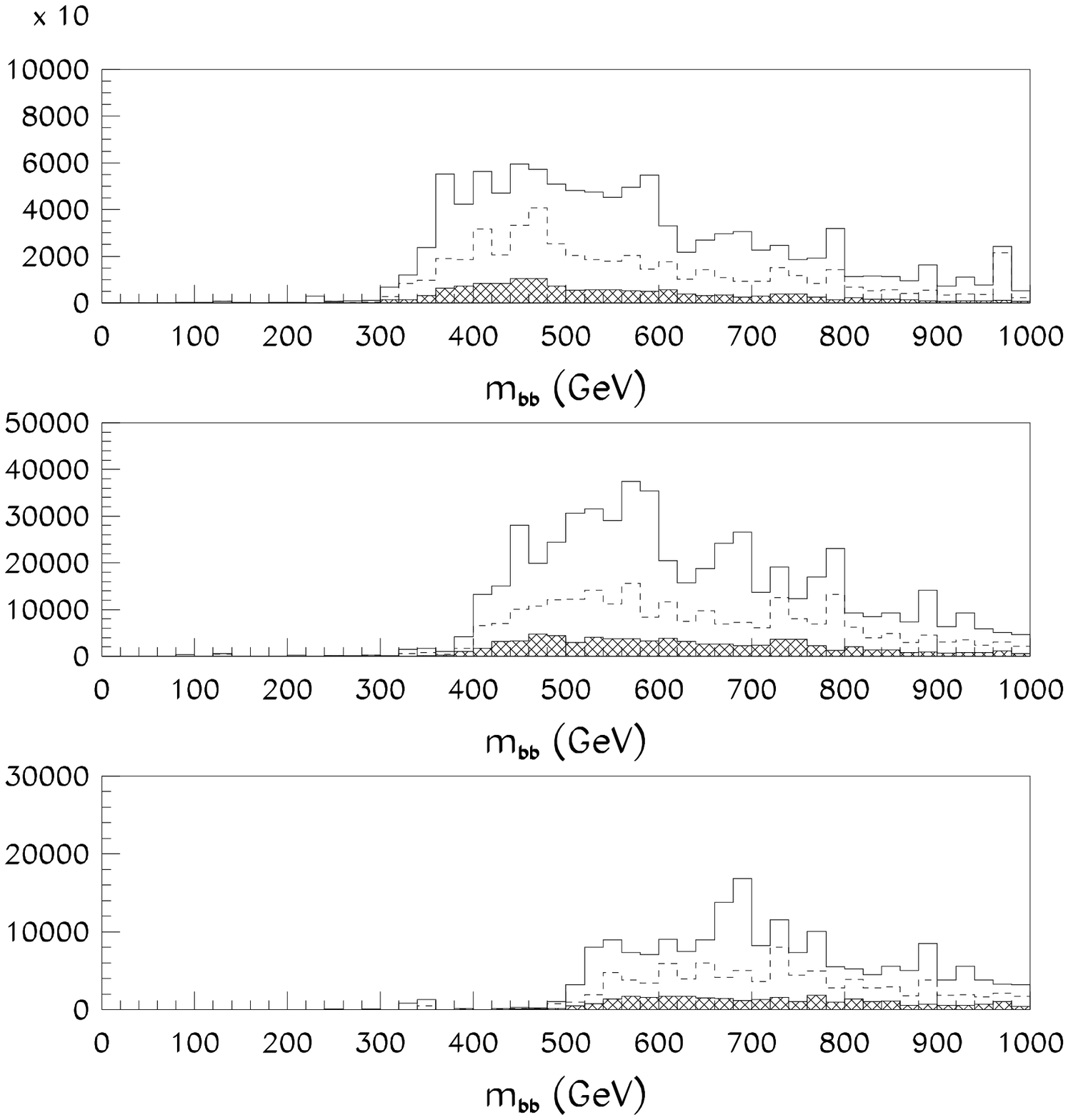,height=16cm,width=16cm}
\end{center}
\caption{\em 
Distribution of the reconstructed \mbb\ mass for background events after 
applying the selection cuts~$\mathrm{S_{500}}$~(top),
$\mathrm{S_{700}}$~(middle) and~$\mathrm{S_{900}}$~(bottom) and the b-tagging procedure, 
for an integrated luminosity of~\lumc\ (ATLAS). The separate contributions from
direct $\mathrm{b \bar b}$~(hashed), from~$\mathrm{b \bar b~+~gb}$~(dashed) and the total spectrum
(solid) are also shown.
\label{FHIGGS:backgb}}
\end{Fighere}

 \subsection{Expected signal significances} 

Table~\ref{TSIGN:Hbblow} (for an integrated luminosity of~\luma) 
and Table~\ref{TSIGN:Hbbhigh} (for an integrated luminosity of~\lumc) show the 
final signal and background rates and expected significances for ATLAS experiment 
as a function 
of the selection procedure adopted, for $\mathrm{\tan \beta~=~30}$ and for 
four values of the Higgs-boson masses. The selection criteria ($\mathrm{S_{300}}$ 
to~$\mathrm{S_{900}}$), as well as the mass window for the reconstructed~\mbb, were 
optimised separately for each mass (a mass window of~$\mathrm{\sim~\pm 2\sigma_m}$ 
was used). 

The overall conclusion is that, even for an integrated luminosity of~\lumc,
the extraction of the signal for $\mathrm{b \bar b H/A}$ production with 
$\mathrm{H/A \to b \bar b}$ decay will be very difficult. Although the expected rates
and significances are higher for a selection requiring only three 
identified b-jets, it is unlikely that the systematic uncertainties on the
background shape can be controlled to the required precision in this case,
where the signal-to-background ratio is lowest~(S/B~$\sim$~1\%).
A selection requiring at least four identified b-jets yields a more favourable 
signal-to-background ratio of~$\sim$~3\% and also a more favourable ratio
(of~$\sim$~67\%) of irreducible-to-total 
background\footnote{The irreducible background consists
of the events containing the required number of identified true b-jets.}. As
mentioned previously, this latter ratio can be easily extracted from the 
numbers given in~Table~\ref{TBDG:efficiency2} for~R~=~$10^6$ and~R~=~$10^2$.

For a selection requiring four identified b-jets and for an integrated 
luminosity of~\lumc\ (ATLAS), a nominal significance larger than~$\mathrm{5\sigma}$ could be 
achieved for $\mathrm{\tan \beta~>}$~29 (\mH~=~500~GeV) and for $\tan \beta~>$~20
(\mH~=~300~GeV). This coverage in the parameter space can be compared with that
of the $\tau \tau$~channel, which extends down to~$\mathrm{\tan \beta~>~25}$ 
(\mH~=~500~GeV) and to~$\mathrm{\tan \beta~>~8}$ (\mH~=~300~GeV).
For an integrated luminosity of \luma, the respective lower limits
on~$\tan \beta$ are $\mathrm{\tan \beta~>~42}$ and $\mathrm{\tan \beta~>~30}$ for the
$\mathrm{b \bar b}$~channel, and $\mathrm{\tan \beta~>~33}$ and $\mathrm{\tan \beta~>~11}$ for 
the $\mathrm{\tau \tau}$~channel.

It should be stressed again that the sensitivities quoted for the
$\mathrm{b \bar b}$~channel are rather on the optimistic side, since the estimates
of the QCD~background are very uncertain (the PYTHIA results could be 
under-estimated by a factor of~3), and the
assumptions used for the b-tagging performance are more optimistic than
the results from recent work for the ATLAS Inner Detector 
Performance~TDR~\cite{ID-TDR}. In addition, systematic uncertainties due
to the lack of knowledge of the background shape have not been taken into 
account in the significance estimates of Tables~\ref{TSIGN:Hbblow} 
and~\ref{TSIGN:Hbbhigh}.

\newpage
                                  \begin{Tabhere}
      \newcommand{\lstrut}{{$\strut\atop\strut$}}
             \caption {\em Expected signal and background rates inside the 
                        \mbb~mass window, for four values of~\mH\ and~\mA, 
                        for $\mathrm{\tan \beta~=~30}$ and for an integrated luminosity 
                        of~\luma\ (ATLAS). Also shown are the expected 
                        signal-to-background ratios and signal significances
                        for various selection algorithms (see text).
                       \label{TSIGN:Hbblow}}
                                     \vspace{2mm}
                                   \begin{center}
       \begin{tabular}{|c||c|c|c||c|}
                                    \hline 
 $\mathrm{b \bar b H/A}$ with $\mathrm{H/A \to b \bar b}$& Signal (S)&
 Background (B) & S/B & $\mathrm{S/\sqrt B}$ \\ \hline \hline
\cline{2-5} \multicolumn{1}{|c||}{}& \multicolumn{4}{|c|}{ \mH~=~300~GeV } \\
\hline
Selection $\mathrm{S_{300}}$ &  &  &  & \\ \hline \hline
 Three b-jets  & 3200 & $2.1 \cdot 10^5$ & 1.5\% & 7.0 \\  \hline
 Algorithm A   &  630 & $2.4 \cdot 10^4$ & 2.6\% & 4.1 \\  \hline
 Algorithm B   & 1075 & $4.6 \cdot 10^4$ & 2.3\% & 5.0 \\  \hline
\hline 
\cline{2-5} \multicolumn{1}{|c||}{}& \multicolumn{4}{|c|}{ \mH~=~500~GeV } \\
\hline \hline
Selection $\mathrm{S_{500}}$ &  &  &  & \\ \hline
 Three b-jets  &  855 & $5.9 \cdot 10^4$ & 1.4\% & 3.5 \\  \hline
 Algorithm A   &  200 & $6.1 \cdot 10^3$ & 3.3\% & 2.6 \\  \hline
 Algorithm B   &  265 & $1.1 \cdot 10^4$ & 2.4\% & 2.5 \\  \hline
\hline 
\cline{2-5} \multicolumn{1}{|c||}{}& \multicolumn{4}{|c|}{ \mH~=~700~GeV } \\
\hline \hline
Selection $\mathrm{S_{700}}$ &  &  &  & \\ \hline
 Three b-jets  &  260 & $2.6 \cdot 10^4$ & 1.0\% & 1.6 \\  \hline
 Algorithm A   &   50 & $1.8 \cdot 10^3$ & 2.8\% & 1.2 \\  \hline
 Algorithm B   &   70 & $5.1 \cdot 10^3$ & 1.4\% & 1.0 \\  \hline
\hline 
\cline{2-5} \multicolumn{1}{|c||}{}& \multicolumn{4}{|c|}{ \mH~=~900~GeV } \\
\hline \hline
 Selection $\mathrm{S_{900}}$ &  &  &  & \\ \hline
 Three b-jets  &   90 & $7.3 \cdot 10^3$ & 1.2\% & 1.0 \\  \hline
 Algorithm A   &   16 & $5.2 \cdot 10^2$ & 3.0\% & 0.7 \\ \hline
 Algorithm B   &   25 & $9.9 \cdot 10^2$ & 2.5\% & 0.8 \\ \hline 
\end{tabular}                                                                   
\end{center}                                                                    
                                    \end{Tabhere}

\newpage

                                  \begin{Tabhere}
      \newcommand{\lstrut}{{$\strut\atop\strut$}}
             \caption {\em Expected signal and background rates inside the 
                        \mbb~mass window, for four values of~\mH\ and~\mA, 
                        for $\mathrm{\tan \beta~=~30}$ and for an integrated luminosity 
                        of~\lumc\ (ATLAS). Also shown are the expected 
                        signal-to-background ratios and signal significances
                        for various selection algorithms (see text).
                       \label{TSIGN:Hbbhigh}}
                                     \vspace{2mm}
                                   \begin{center}
       \begin{tabular}{|c||c|c|c||c|}
                                    \hline 
 $\mathrm{b \bar b H/A}$ with $\mathrm{H/A \to b \bar b}$ & Signal (S)& Background (B) & S/B &
 $\mathrm{S/\sqrt B}$ \\ \hline \hline
\cline{2-5} \multicolumn{1}{|c||}{}& \multicolumn{4}{|c|}{ \mH~=~300~GeV } \\
\hline
Selection $\mathrm{S_{300}}$ &  &  &  & \\ \hline
 Three b-jets  & 18700 & $1.4 \cdot 10^6$ & 1.3\% & 15.8 \\  \hline
 Algorithm A   &  3080 & $1.3 \cdot 10^5$ & 2.4\% &  8.5 \\  \hline
 Algorithm B   &  5270 & $2.4 \cdot 10^5$ & 2.2\% & 10.7 \\  \hline
\hline 
\cline{2-5} \multicolumn{1}{|c||}{}& \multicolumn{4}{|c|}{ \mH~=~500~GeV } \\
\hline \hline
Selection $\mathrm{S_{500}}$ &  &  &  & \\ \hline
 Three b-jets  &  5000 & $3.9 \cdot 10^5$ & 1.3\% & 8.0 \\  \hline
 Algorithm A   &   960 & $3.3 \cdot 10^4$ & 2.9\% & 5.3 \\  \hline
 Algorithm B   &  1300 & $6.0 \cdot 10^4$ & 2.2\% & 5.3 \\  \hline
\hline 
\cline{2-5} \multicolumn{1}{|c||}{}& \multicolumn{4}{|c|}{ \mH~=~700~GeV } \\
\hline \hline
Selection $\mathrm{S_{900}}$ &  &  &  & \\ \hline
 Three b-jets  &  1530 & $1.7 \cdot 10^5$ & 0.9\% & 3.7 \\  \hline
 Algorithm A   &   240 & $1.0 \cdot 10^4$ & 2.4\% & 2.4 \\  \hline
 Algorithm B   &   350 & $2.7 \cdot 10^4$ & 1.3\% & 2.1 \\  \hline
\hline 
\cline{2-5} \multicolumn{1}{|c||}{}& \multicolumn{4}{|c|}{ \mH~=~900~GeV } \\
\hline \hline
Selection $\mathrm{S_{900}}$ &  &  &  & \\ \hline
 Three b-jets  &   524 & $4.8 \cdot 10^4$ & 1.1\% & 2.4 \\  \hline
 Algorithm A   &    80 & $2.8 \cdot 10^3$ & 2.8\% & 1.5 \\  \hline
 Algorithm B   &   122 & $5.2 \cdot 10^3$ & 2.3\% & 1.7 \\  \hline 
\end{tabular}                                                                   
\end{center}            
                                                        
                                    \end{Tabhere}
\newpage

\boldmath
\section{Observability of $\mathrm{H \to hh \to b \bar b b \bar b}$ \\
  for small $\mathrm{\tan \beta}$}
\unboldmath

\subsection{ Signal events}

Two points in the MSSM parameter space were chosen for a detailed study
of this channel: \mH~=~300~GeV with $\mathrm{\tan \beta}$~=~1.5 and $\mathrm{\tan \beta}$~=~3.0,
corresponding respectively to \mh~=~77.8~GeV and \mh~=~98~GeV 
(2-loop calculation). The final state in this channel can be fully
reconstructed with good mass resolution through the use of the constraint,
\mbb~=~\mh. Four identified b-jets are required with two 
$b \bar b$~combinations reconstructed with \mbb\ close to~\mh. 
As shown also in~\cite{Note-043}, with the expected ATLAS calorimeter
performance, a signal acceptance of~70\%~to~80\% can be achieved using a mass 
window, \mbb~=~\mh~$\pm$~25~GeV, for the h-boson reconstruction and, after 
applying the constraint on~\mh, a mass window, \mbbbb~=~\mH~$\pm$~20~GeV.

For the signal events, the jet transverse-energy spectrum is rather hard, 
with $\mathrm{<\pT^{j_1}>}$~=~100~GeV and $\mathrm{<\pT^{j_4}>}$~=~36~GeV for jets within
the Inner Detector acceptance for b-tagging. For the high-luminosity case,
at least four jets reconstructed with \pT~$>$~40~GeV (before energy recalibration) 
are required, yielding
an acceptance of~$\sim$~25\% for the signal events. 
Table~\ref{THIGGS:efficiency3} shows the expected signal rates
for two different jet-selection algorithms described below:
\begin{itemize}
\item Algorithm A:

 $\bullet$ the four most energetic jets are required to be identified as b-jets;

 $\bullet$ all possible $\mathrm{b \bar b}$~combinations are then reconstructed;
   
 $\bullet$ the best pair of combinations for the reconstruction of~\mh\ is 
 chosen by minimising 
 $\mathrm{\chi^2~=~(\mbb_{,1}~-~\mh)^2~+~(\mbb_{,2}~-~\mh)^2}$;
 
 $\bullet$ both pairs are required to satisfy \mbb~=~\mh~$\pm$~25~GeV;
 
 $\bullet$ the event is accepted if \mbbbb~=~\mH~$\pm$~20~GeV after applying 
           a constraint on~\mh.
\item Algorithm B:

 $\bullet$ all possible $\mathrm{jj}$ combinations inside the Inner Detector acceptance 
  are considered;
      
 $\bullet$ the best pair of combinations for the reconstruction of~\mh\ is 
 chosen by minimising 
 $\mathrm{\chi^2~=~(\mjj_{,1}~-~\mh)^2~+~(\mjj_{,2}~-~\mh)^2}$;

 $\bullet$ all four jets chosen in this way are required to be identified as
           b-jets;
 
 $\bullet$ both pairs are required to satisfy \mbb~=~\mh~$\pm$~25~GeV;
 
 $\bullet$ the event is accepted if \mbbbb~=~\mH~$\pm$~20~GeV after applying 
           a constraint on~\mh.
\end{itemize}
A careful comparison of these algorithms with other possible selection
methods has shown that Algorithm B is close to optimal.

                                 \begin{Tabhere}
      \newcommand{\lstrut}{{$\strut\atop\strut$}}
             \caption {\em Numbers of expected $\mathrm{H \to hh \to b\bar b b \bar b}$
                        signal events as a function of the selection algorithm,
                        for an integrated luminosity of~\lumc\ (ATLAS). The 
                        high-luminosity b-tagging performance is assumed
                        ($\mathrm{\epsilon_b~=~50\%}$, $\mathrm{\epsilon_c~=~10\%}$
                         and $\mathrm{R~=~100}$).
                       \label{THIGGS:efficiency3}}
                                     \vspace{2mm}
                                   \begin{center}
       \begin{tabular}{|c||c||c|c|c|c|}
                                    \hline 

               &          &      Four     &    Four    & + \mbb\ within & 
 +~\mH\ within    \\ 
 \mH~=~300~GeV & $\sigma$ & reconstructed & identified & $\pm$~25~GeV   & 
 $\pm$~20~GeV     \\ 
               &  (pb)    &     jets      &   b-jets   &    of~\mh\    & 
 of~\mH.         \\ 
                                    \hline \hline 
\cline{4-6} \multicolumn{3}{|c|}{ }& \multicolumn{3}{|c|}{ Algorithm A } \\
\hline
  \tanb~=~3.0 &  0.76  & $5.8 \cdot 10^4$  & 350  & 154  & 125 \\
  \tanb~=~1.5 &  1.73  & $1.6 \cdot 10^5$   & 800  & 510  & 430 \\
\hline 
\cline{4-6} \multicolumn{3}{|c|}{ }& \multicolumn{3}{|c|}{ Algorithm B } \\
                                    \hline \hline 
  \tanb~=~3.0 &  0.76  & $5.8 \cdot 10^4$  & 640   &  470 &  360 \\
  \tanb~=~1.5 &  1.73  & $1.6 \cdot 10^5$  & 1760  & 1560 &  1360 \\
\hline 
\end{tabular}                                                                   
\end{center}                                                                    
                                    \end{Tabhere}

\subsection{Background events}

As in Section~2.2, the background events arise dominantly from QCD multi-jet 
production.
Table~\ref{TBDG:efficiency4} shows the expected rates of reconstructed
 multi-jet background events for the various QCD sub-processes and for an
 integrated luminosity of \lumc. This Table also shows the contributions from
events containing one or more true b-jets to each of the sub-processes; 
for the dominant gg sub-process, only 0.01\% of the events contain four
true b-jets in the final state, whereas this factor increases to 0.8\% for
the direct $\mathrm{b \bar b}$ sub-process. Nevertheless, since the production
cross-sections differ by two orders of magnitude, the direct $\mathrm{b \bar b}$ production
process contributes only $\sim$ 10\% of the total QCD background after having
required four tagged b-jets, as shown in Table~\ref{TBGD:efficiency5}.
The dominant background is from the gb sub-process, and the irreducible
background from four true b-jets is shown to amount to $\sim$ 70\% of the
total multi-jet background. This can be deduced from a comparison of the expected
background rates in Table~\ref{TBGD:efficiency5} for $\mathrm{R~=~10^6}$ 
(almost infinite rejection of non-b-jets) and for $\mathrm{R~=~10^2}$ 
(default b-tagging performance). This is also illustrated in
 Fig.~\ref{FHIGGS:mbbbb98log}, which displays the mass distributions expected 
 for the candidate $\mathrm{h \to b \bar b}$ decays for the three dominant sub-processes 
 and for the events accepted by algorithm B. Finally, 
Table~\ref{THIGGS:efficiency6} shows the expected rates of background events at
 each step of the selection procedures described as Algorithms A and B.

\newpage
                                  \begin{Tabhere}
      \newcommand{\lstrut}{{$\strut\atop\strut$}}
             \caption {\em Expected rates of QCD multi-jet background events
                           for different sub-processes and for an integrated
                           luminosity of \lumc\ (ATLAS). Also shown are the contributions
                           to the total background from events 
                           containing one or more true b-jets.
                       \label{TBDG:efficiency4}}
                                     \vspace{2mm}
                                   \begin{center}
       \begin{tabular}{|c|c||c|c|c|c|}
                                    \hline 
 Sub-process &   Four jets         & Jet1  & + jet2  & + jet3  & + jet4  \\
             & with \pT~$>$~40~GeV &  = true  & = true     & = true     & = true     \\ 
             & and $|\eta|~<~2.5$  &  b-jet & b-jet    & b-jet    & b-jet    \\ 
                                    \hline \hline
$\mathrm{gg, q \bar q \to b \bar b}$&$ 2.0 \cdot 10^8$   & 70\% & 30\%  & 2\%    & 0.8\%
\\
$\mathrm{gb \to gb}$                &$ 1.2 \cdot 10^9$   & 45\% & 10\%  & 1\%    & 0.3\%
\\
$\mathrm{gg,qq \to gg,qq}$          &$ 1.7 \cdot 10^{10}$&  3\% & 0.4\% & 0.04\% & 0.01\%
\\
\hline
\end{tabular}                                                                   
\end{center}                                                                    
                                    \end{Tabhere}

                                  \begin{Tabhere}
      \newcommand{\lstrut}{{$\strut\atop\strut$}}
             \caption {\em Expected rates of reconstructed background events 
                        after jet-selection cuts as a function of the 
                        b-tagging performance, for an integrated 
                        luminosity of~\lumc\ (ATLAS).
                       \label{TBGD:efficiency5}}
                                     \vspace{2mm}
                                   \begin{center}
       \begin{tabular}{|c||c|c|c|c||c|}
                                    \hline 
 Sub-process &     Four jets       & Jet1  & + jet2  & + jet3  & +jet4    \\
             & with \pT~$>$~40~GeV & tagged &  tagged  &  tagged  & tagged    \\
             &  and $\mathrm{|\eta|<2.5}$   & as b-jet  &  as b-jet  &  as  b-jet  & as b-jet     \\
                                    \hline \hline
\cline{3-6} \multicolumn{2}{|c|}{}& \multicolumn{4}{|c|}{ $\epsilon_b~=~0.5$,
$ \mathrm{\epsilon_c~=~10^{-6}}$, \ \ R~=~10$^6$ } \\
\hline \hline
$\mathrm{gg, q \bar q \to b \bar b}$   &
   $2.0 \cdot 10^8$ & $6.9 \cdot 10^7$ & $1.5 \cdot 10^7$ & $5.0 \cdot 10^5$ & $9.7 \cdot 10^4$ \\
$\mathrm{gb \to gb}$   &
   $1.2 \cdot 10^9$ & $2.7 \cdot 10^8$ & $2.9 \cdot 10^7$ & $1.5 \cdot 10^6$ & $2.1 \cdot 10^5$ \\
$\mathrm{gg,qq \to gg,qq}$   & 
 $1.7 \cdot 10^{10}$  & $2.6 \cdot 10^8$ & $1.6 \cdot 10^7$ & $8.6 \cdot 10^5$ & $1.2 \cdot 10^5$ \\ \hline
Total &  
 $1.8 \cdot 10^{10}$  & $6.0 \cdot 10^8$ & $6.0 \cdot 10^7$ & $2.9 \cdot 10^6$ & $4.3 \cdot 10^5$ \\
\hline \hline
\cline{3-6} \multicolumn{2}{|c|}{}& \multicolumn{4}{|c|}{ $\mathrm{\epsilon_b~=~0.5}$,
$ \mathrm{\epsilon_c~=~0.1}$, \ \ R~=~$10^2$ } \\
                                    \hline \hline
$\mathrm{gg, q \bar q \to b \bar b}$   &
 $2.0 \cdot 10^8$    & $7.0 \cdot 10^7$  & $1.6 \cdot 10^7$  & $9.5 \cdot 10^5$ &  $1.2 \cdot 10^5$ \\
$\mathrm{gb \to gb}$   &
 $1.2 \cdot 10^9$    & $2.8 \cdot 10^8$  & $3.4 \cdot 10^7$  & $2.6 \cdot 10^6$ &  $2.9 \cdot 10^5$ \\
$\mathrm{gg,qq \to gg,qq}$  & 
 $1.7 \cdot 10^{10}$ & $5.2 \cdot 10^8$  & $2.8 \cdot 10^7$  & $2.1 \cdot 10^6$ &  $1.9 \cdot 10^5$ \\ \hline
Total &  
 $1.8 \cdot 10^{10}$ & $8.7 \cdot 10^8$  & $7.8 \cdot 10^7$  & $5.6 \cdot 10^6$ &  $6.0 \cdot 10^5$ \\
\hline \hline
\cline{3-6} \multicolumn{2}{|c|}{}& \multicolumn{4}{|c|}{ $\epsilon_b~=~0.6$,
$ \mathrm{\epsilon_c~=~0.1}$, \ \ R~=~$10^2$ } \\
                                    \hline \hline
$\mathrm{gg, q \bar q \to b \bar b}$   &
 $2.0 \cdot 10^8$    & $8.4 \cdot 10^7$  &  $2.2 \cdot 10^7$  & $1.5 \cdot 10^6$ &  $2.4 \cdot 10^5$ \\
$\mathrm{gb \to gb}$   &
 $1.2 \cdot 10^9$    & $3.3 \cdot 10^8$  &  $4.8 \cdot 10^7$  & $4.2 \cdot 10^6$ &  $5.6 \cdot 10^5$ \\
$\mathrm{gg,qq \to gg,qq}$  & 
 $1.7 \cdot 10^{10}$ & $5.7 \cdot 10^8$  &  $3.6 \cdot 10^7$  & $3.2 \cdot 10^6$ &  $3.5 \cdot 10^5$ \\ \hline
Total &  
 $1.8 \cdot 10^{10}$ & $9.8 \cdot 10^8$  &  $1.1 \cdot 10^8$  & $8.9 \cdot 10^6$ &  $1.1 \cdot 10^6$\\
\hline
\end{tabular}                                                                   
\end{center}                                                                    
                                    \end{Tabhere}

                                     \vspace{-1cm}
\begin{Fighere}
\begin{center}
\epsfig{file=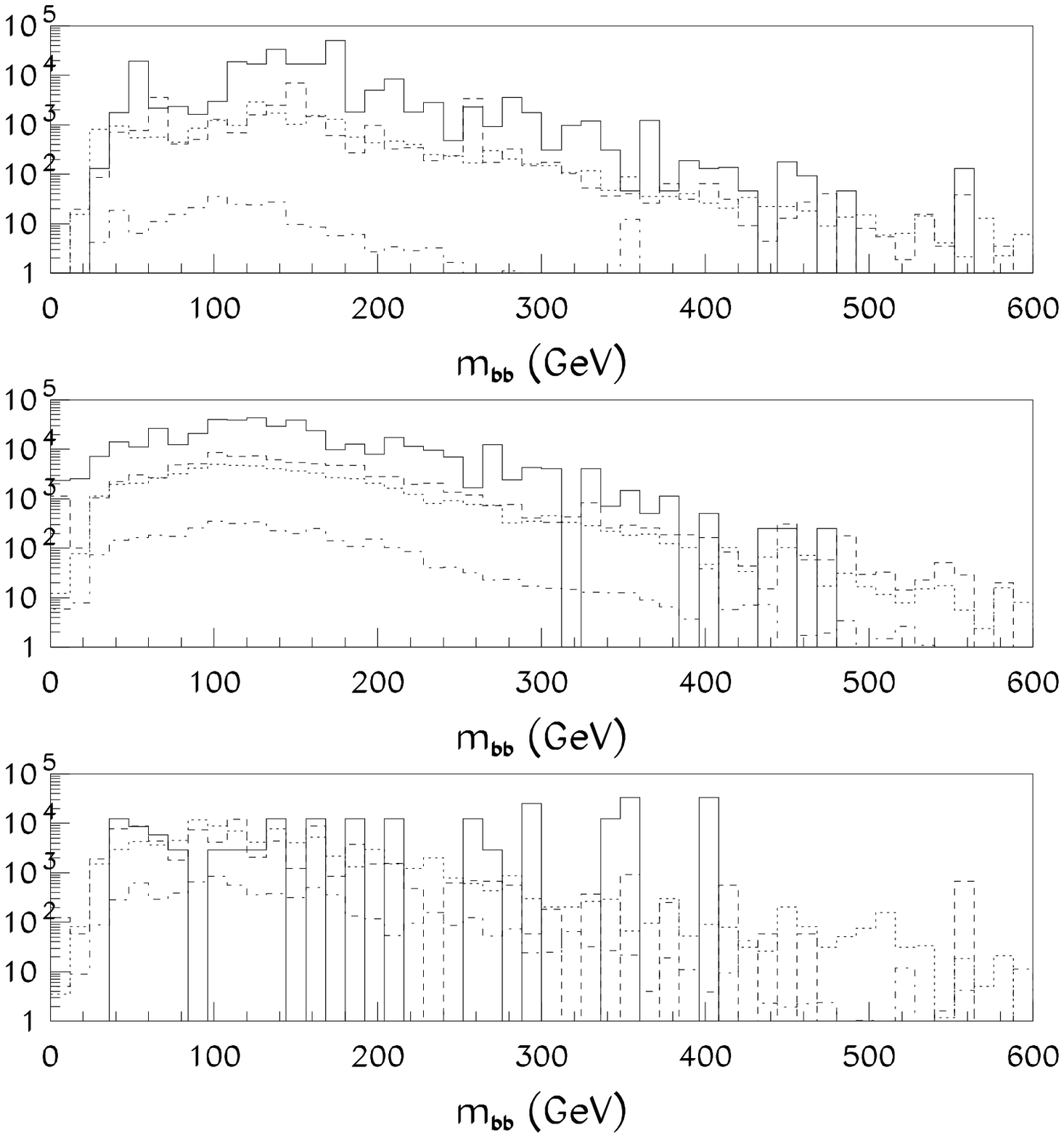,height=8cm,width=10cm}
\end{center}
\caption{\em 
          Distribution of the reconstructed \mbb\ for background events 
          after applying selection~B, shown separately for the various 
          sub-processes: $\mathrm{b \bar b}$~(top), gb~(middle) 
          and~$\mathrm{gg+gq+qq}$~(bottom). Also shown separately are the contributions 
          from events containing four true b-jets~(solid), three true 
          b-jets~(dashed), two true b-jets~(dots) and one or less true 
          b-jets~(dot-dashed). 
         \label{FHIGGS:mbbbb98log}}
\end{Fighere}

                                 \begin{Tabhere}
                                     \vspace{-5mm}
      \newcommand{\lstrut}{{$\strut\atop\strut$}}
             \caption {\em Expected rates of reconstructed background events 
                        as a function of the selection cuts, for algorithms~A
                        and~B and  for an integrated luminosity of~\lumc\ (ATLAS). 
                        The default high luminosity b-tagging performance is assumed.
                       \label{THIGGS:efficiency6}}
                                   \begin{center}
       \begin{tabular}{|c||c||c|c|c|c|}
                                    \hline 
Sub-process & $\mathrm{\sigma}$  (mb)       &       Four jets       &  Four jets & 
\mh~within  & \mH~within   \\ 
            & \pT~$>$~10~GeV & with \pT~$>$~40~GeV   & identified & 
$\pm~25$~GeV & $\pm~20 $ GeV \\ 
            &                      &    and $\mathrm{|\eta|<2.5}$   &  as b-jets &
of nominal  & of nominal   \\ 
                                    \hline \hline 
\cline{4-6} \multicolumn{3}{|c|}{}& \multicolumn{3}{|c|}{ Algorithm A} \\ 
\hline \hline
$\mathrm{b \bar b}$  & $2.1 \cdot 10^{-3}$ & $2.0 \cdot 10^{8}$  & $1.2 \cdot 10^5$ & 
 2000  & 200  \\
$gb$        & $7.1 \cdot 10^{-2}$ & $1.2 \cdot 10^{9}$  & $2.8 \cdot 10^5$ &
13000  & 1900 \\
$\mathrm{gg,gq,qq}$  & $ 5.9 $             & $1.7 \cdot 10^{10}$ & $1.9 \cdot 10^5$ & 
 2000  &  500 \\ \hline
  Total      &                    & $1.8 \cdot 10^{10}$ & $5.9 \cdot 10^5$ & 
17000  & 2600 \\
\hline 
\cline{4-6} \multicolumn{3}{|c|}{}& \multicolumn{3}{|c|}{ Algorithm B} \\ 
\hline \hline
$\mathrm{b \bar b}$  & $2.1 \cdot 10^{-3}$ & $2.0 \cdot 10^{8}$  & $1.4 \cdot 10^5$ &
 3000  &  500  \\
$\mathrm{gb}$        & $7.1 \cdot 10^{-2}$ & $1.2 \cdot 10^{9}$  & $2.9 \cdot 10^5$ &
24000  & 2500  \\
$\mathrm{gg,gq,qq}$  & $ 5.9 $             & $1.7 \cdot 10^{10}$ & $1.9 \cdot 10^5$ &
14000  & 1000  \\ \hline
  Total      &                    & $1.8 \cdot 10^{10}$ & $6.2 \cdot 10^5$ & 
41000  & 4000  \\
\hline 
\end{tabular}                                                                   
\end{center}                                                                    
                                    \end{Tabhere}

\subsection{Significance}

Table~\ref{THIGGS:Hhhhigh} shows, for both algorithms A and B, the expected
signal and background rates for \mH~=~300~GeV, for \tanb~=1.5 and~3.0, and 
for an integrated luminosity of \lumc. Algorithm B yields a significantly
better signal-to-background ratio and thereby sensitivity, since 
 b-jets from $\mathrm{H \to hh \to b \bar b b \bar b}$ decays are not very often
the hardest jets in the event.

The $\mathrm{m _{b \bar b}}$ and $\mathrm{m_{b \bar b b \bar b}}$ mass distributions for both the
signal and background events selected by algorithm B are shown in
 Fig.~\ref{FHIGGS:mbb98} and  Figure~\ref{FHIGGS:mbbbb98} before the relevant 
mass cuts are applied.
Fig.~\ref{FHIGGS:distribhigh} shows the distributions of $\mathrm{p_T^{b \bar b}}$, 
$\mathrm{p_T^{b \bar b b \bar b}}$, $\mathrm{p_T^{b_1}}$ and $\mathrm{p_T^{b_4}}$
 before energy recalibration
for the accepted signal and background events. After applying the selection cuts
 of algorithm B, these distributions do not differ sufficiently between signal
 and the background to justify further optimisation of the selection cuts.

A sensitivity similar to that expected for the
 $\mathrm{H \to hh \to b \bar b \gamma \gamma}$ channel (see Fig.~\ref{FMSSM:lhcmB})
 appears to
 be also achievable in this channel, since rate is larger, but this would require
 triggering on four-jet events at high luminosity with $\mathrm{p_T^{jet}>}$~40~GeV and
 $\mathrm{|\eta|~<}$~2.5.
 The expected trigger rates are very uncertain, in the range between
 1.6 and 4.5 kHz for PYTHIA and NJETS respectively (before energy recalibration), and also 
too high to be accepted without further cuts at level-2.
 The possibility of using b-tagging to reduce 
the rate will be investigated and compared with other simpler possibilities 
involving somewhat tighter cuts.

This channel has also been studied in the low-luminosity case to evaluate the possible gain
in sensitivity that could be achieved by lowering the jet $\mathrm{E_T}$-threshold.
The results of this study are shown, for algorithm B, for \tanb~=~3.0 and for an 
 integrated luminosity of \luma, in Table~\ref{THIGGS:Hhhlow} as a function of the
 jet $\mathrm{E_T}$-threshold, which was varied from its minimum realistic value
 (before recalibration) of 15~GeV to the high-luminosity value of 40~GeV.
The background rates decrease much faster than the signal rates and the sensitivity 
improves as the jet $E_T$-threshold is raised. For a threshold of 20~GeV,
 Fig.~\ref{FHIGGS:distriblow}
shows the distributions of $\mathrm{p_T^{b \bar b}}$, 
$\mathrm{p_T^{b \bar b b \bar b}}$, $\mathrm{p_T^{b_1}}$
 and $\mathrm{p_T^{b_4}}$ for the accepted signal and 
background events: requiring a  cut on
 $\mathrm{p_T^{b \bar b}}$, typically  $\mathrm{p_T^{b \bar b}~>}$~60~GeV, would obviously improve
 the signal sensitivity. As an illustration, Table~\ref{THIGGS:Hhhlow} also shows 
the resulting expected signal and background rates as well as the signal-to-background
 ratios and signal significances, for events with  $\mathrm{p_T^{b \bar b}~>}$~60~GeV and with
 $\mathrm{p_T^{b_1}~>}$~80~GeV, as a function of the jet $\mathrm{E_T}$-threshold.

Even with optimised cuts, the sensitivity at low luminosity is weaker than that for the 
$\mathrm{H \to hh \to b \bar b \gamma \gamma}$ channel \cite{Note-074},
 and the trigger requirements become very 
demanding, since an $\mathrm{E_T}$-threshold of about 20 GeV would be desirable.

\newpage

                                 \begin{Tabhere}
      \newcommand{\lstrut}{{$\strut\atop\strut$}}
             \caption {\em For \mH~=~300~GeV and two values of \tanb,
      expected signal and background rates, signal-to-background ratios
      and signal significances, for the two selection algorithms A and B described
      in the text and for an integrated luminosity of  \lumc\ (ATLAS).}
             \label{THIGGS:Hhhhigh}
                                     \vspace{2mm}
                                   \begin{center}
       \begin{tabular}{|c||c||c|c|c|c|}
                                    \hline 
 $\mathrm{H \to hh \to b \bar b b \bar b}$ & $\mathrm{\sigma}$(pb) &  S  & B   & S/B    
& $\mathrm{S/\sqrt{B}}$    \\ \hline 
\cline{2-6} \multicolumn{2}{|c||}{}& \multicolumn{4}{|c|}{ Algorithm A} \\ 
                                    \hline \hline 
  \tanb~=~3.0 &  0.76  & 125   & 2600     &  4.8 \% &  2.4  \\
  \tanb~=~1.5 &  1.73  & 430   & 2600     &  16.5 \% &  8.4  \\
\hline 
\cline{2-6} \multicolumn{2}{|c||}{}& \multicolumn{4}{|c|}{ Algorithm B} \\ 
                                    \hline \hline 
  \tanb~=~3.0 &  0.76  & 360    & 4000     & 9.0 \%  & 5.7   \\
  \tanb~=~1.5 &  1.73  & 1360   & 4000     & 34.0 \%  & 21.5  \\
\hline 
\end{tabular}                                                                   
\end{center}                                                                    
                                    \end{Tabhere}

                                 \begin{Tabhere}
      \newcommand{\lstrut}{{$\strut\atop\strut$}}
             \caption {\em For \mH~=~300~GeV and \tanb~=~3.0, expected signal and
                       background rates, signal-to-background ratios and signal
                       significances as a function of the chosen jet $\mathrm{E_T}$-threshold,
                       for an integrated luminosity of  \luma\ (ATLAS).}
             \label{THIGGS:Hhhlow}
                                     \vspace{2mm}
                                   \begin{center}
       \begin{tabular}{|c||c|c|c|c|}
                                    \hline 
 $\mathrm{H \to hh \to b \bar b b \bar b}$ &  S  & B   & S/B    & $\mathrm{S/\sqrt{B}}$    \\ 
     &   &    &     &     \\ \hline 
\cline{2-5} \multicolumn{1}{|c||}{ Jet $\mathrm{E_T}$ threshold }& \multicolumn{4}{|c|}{ Algorithm B} \\ 
                                    \hline \hline 
  15 GeV & 275   & 64 000    &  0.4  \% &  1.1 \\
  20 GeV & 231   & 48 000    &  0.5  \% &  1.1 \\
  30 GeV & 132   & 10 000    &  1.3  \% &  1.3 \\
  40 GeV &  50   &    800    &  6.2  \% &  1.8  \\ \hline
\cline{2-5} \multicolumn{1}{|c||}{}& \multicolumn{4}{|c|}
{ $\mathrm{+ p_T^{bb}>60}$ GeV and $\mathrm{p_T^{b_1}>80}$ GeV } \\ 
                                    \hline \hline 
  15 GeV & 160   & 3 900  &  4.1  \% & 2.6  \\
  20 GeV & 132   & 2 500  &  5.3  \% & 2.6  \\
  30 GeV &  75   & 1 500  &  5.0  \% & 1.9  \\
  40 GeV &  30   &   400  &  7.5  \% & 1.5  \\
 \hline 
\end{tabular}                                                                   
\end{center}                                                                    
                                    \end{Tabhere}

\newpage

\begin{Fighere}
\begin{center}
\epsfig{file=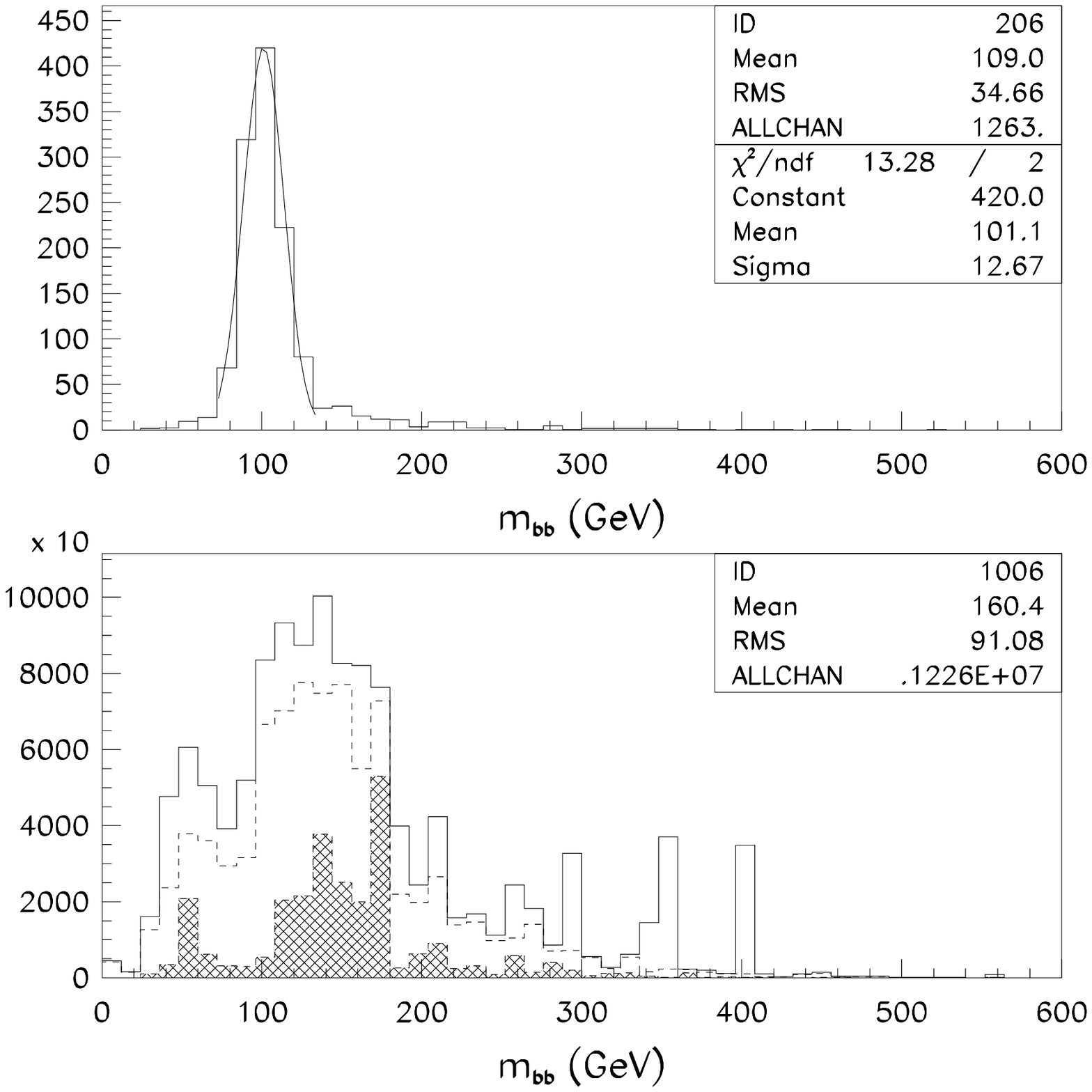,height=14cm,width=14cm}
\end{center}
\caption{\em
Distribution of the reconstructed $\mathrm{m_{b \bar b}}$ mass for the two best combinations
in events containing four b-jets selected by algorithm B and for an integrated
luminosity of \lumc\ (ATLAS). The top distribution is for $\mathrm{H \to hh \to b \bar b b \bar b}$
 signal events with $\mathrm{m_{H}}$~=~300~GeV and \tanb~=~1.5, corresponding to $\mathrm{m_h}$~=~98~GeV.
The bottom distribution is for the background from multi-jet production (solid), where the 
contributions from direct $\mathrm{b \bar b}$ (cross-hatched) and $\mathrm{b \bar b+ gb}$ production (dashed)
are also shown.
\label{FHIGGS:mbb98}}
\end{Fighere}

\newpage
\begin{Fighere}
\begin{center}     
\epsfig{file=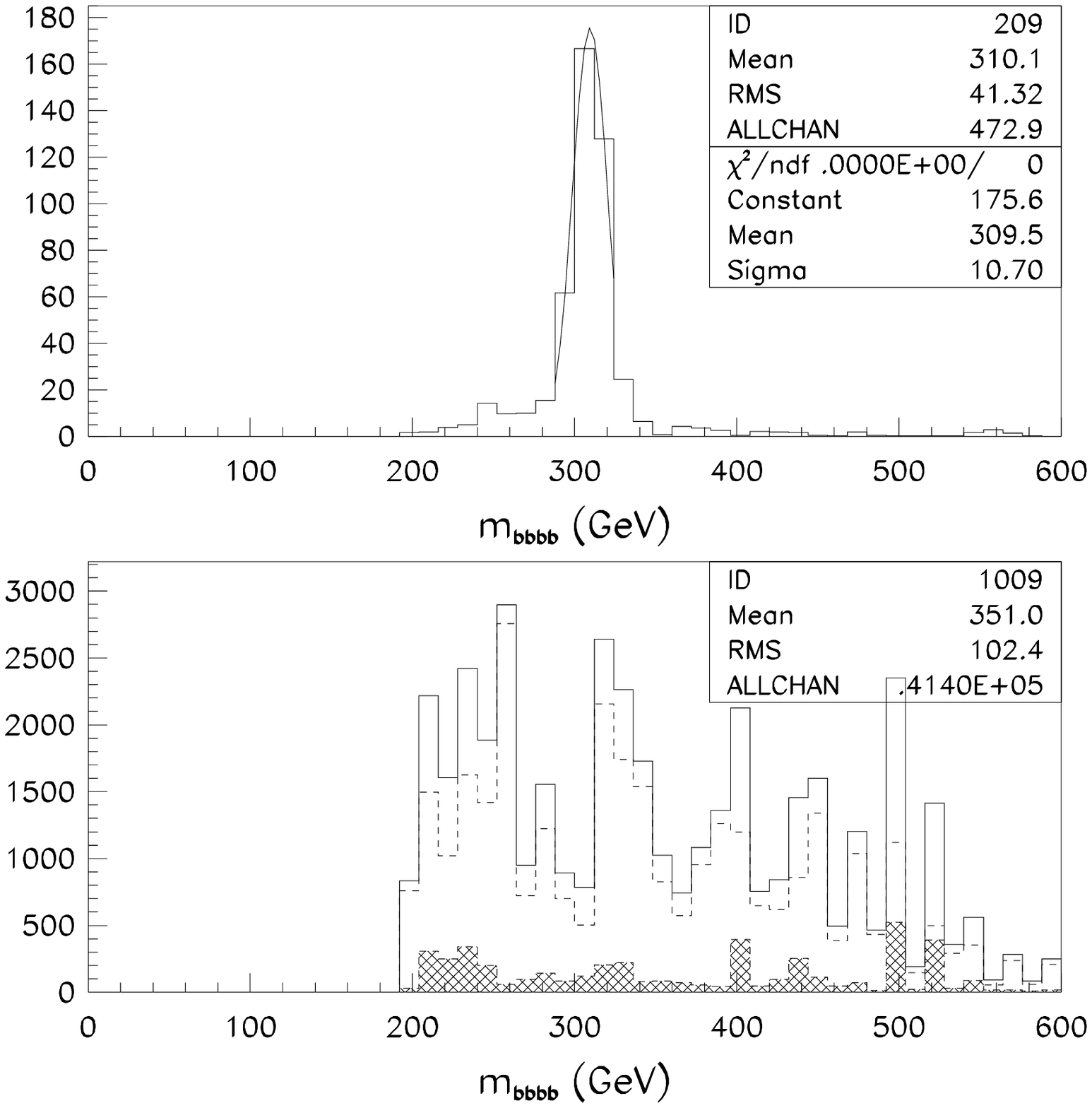,height=14cm,width=14cm}
\end{center}
\caption{\em 
Same as Fig.~\ref{FHIGGS:mbb98} for the $\mathrm{m_{b \bar b b \bar b}}$ spectrum after applying 
a constraint, $\mathrm{m_{b \bar b}~=~m_h}$, on the two best $b \bar b$ combinations.
\label{FHIGGS:mbbbb98}}
\end{Fighere}

\newpage
\begin{Fighere}
\begin{center}     
\epsfig{file=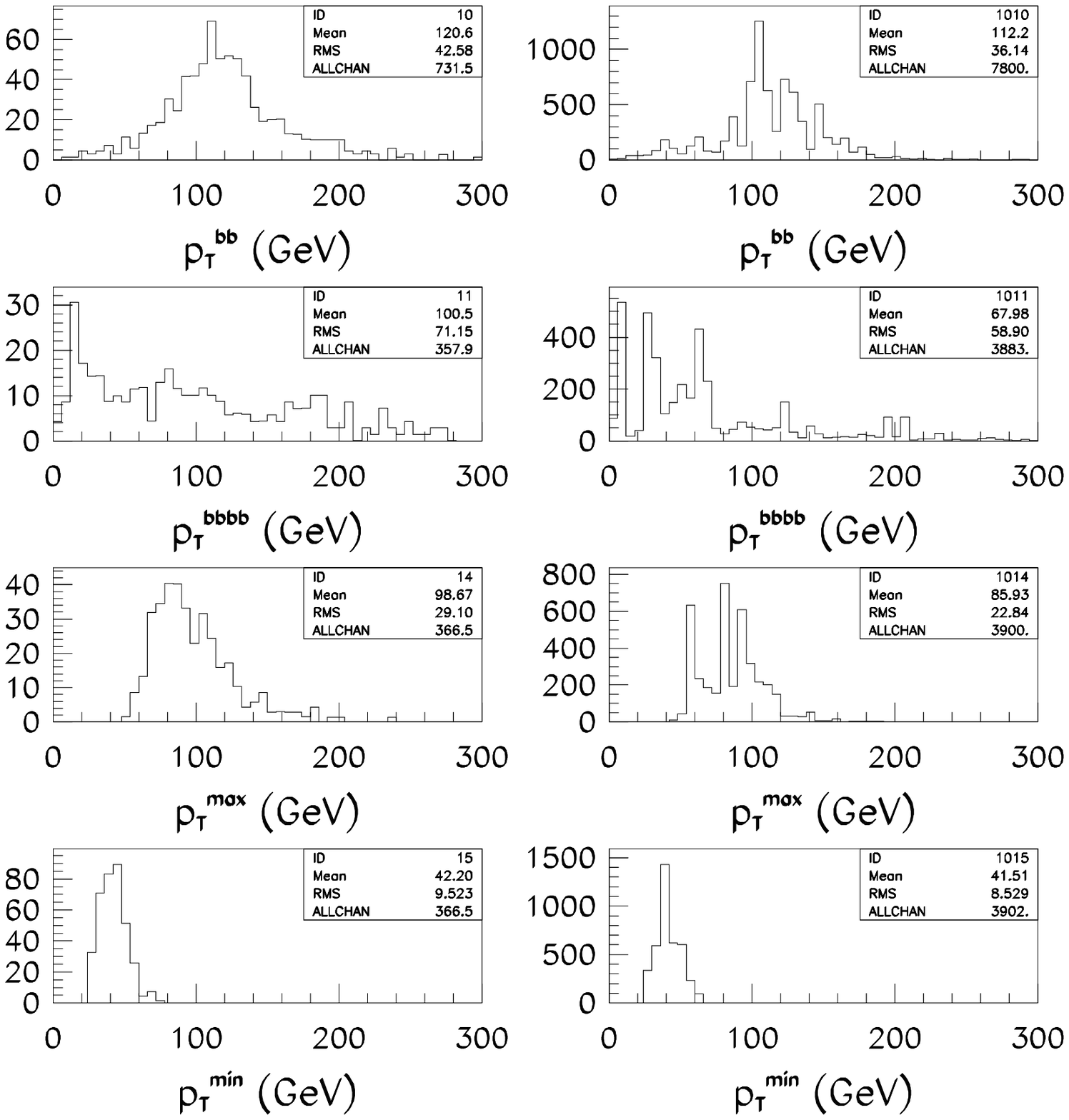,height=16cm,width=10cm}
\end{center}
\caption{\em
 Distributions of $\mathrm{p_T^{b \bar b}}$, $\mathrm{p_T^{b \bar b b \bar b}}$,
 $\mathrm{p_T^{b_1}}$ and $\mathrm{p_T^{b_4}}$
 for signal (left) and background (right) after applying the selection B with 
 a jet $\mathrm{E_T}$-threshold of  40 GeV, for an integrated luminosity of \lumc\ (ATLAS).
\label{FHIGGS:distribhigh}}
\end{Fighere}

\begin{Fighere}
\begin{center}     
\epsfig{file=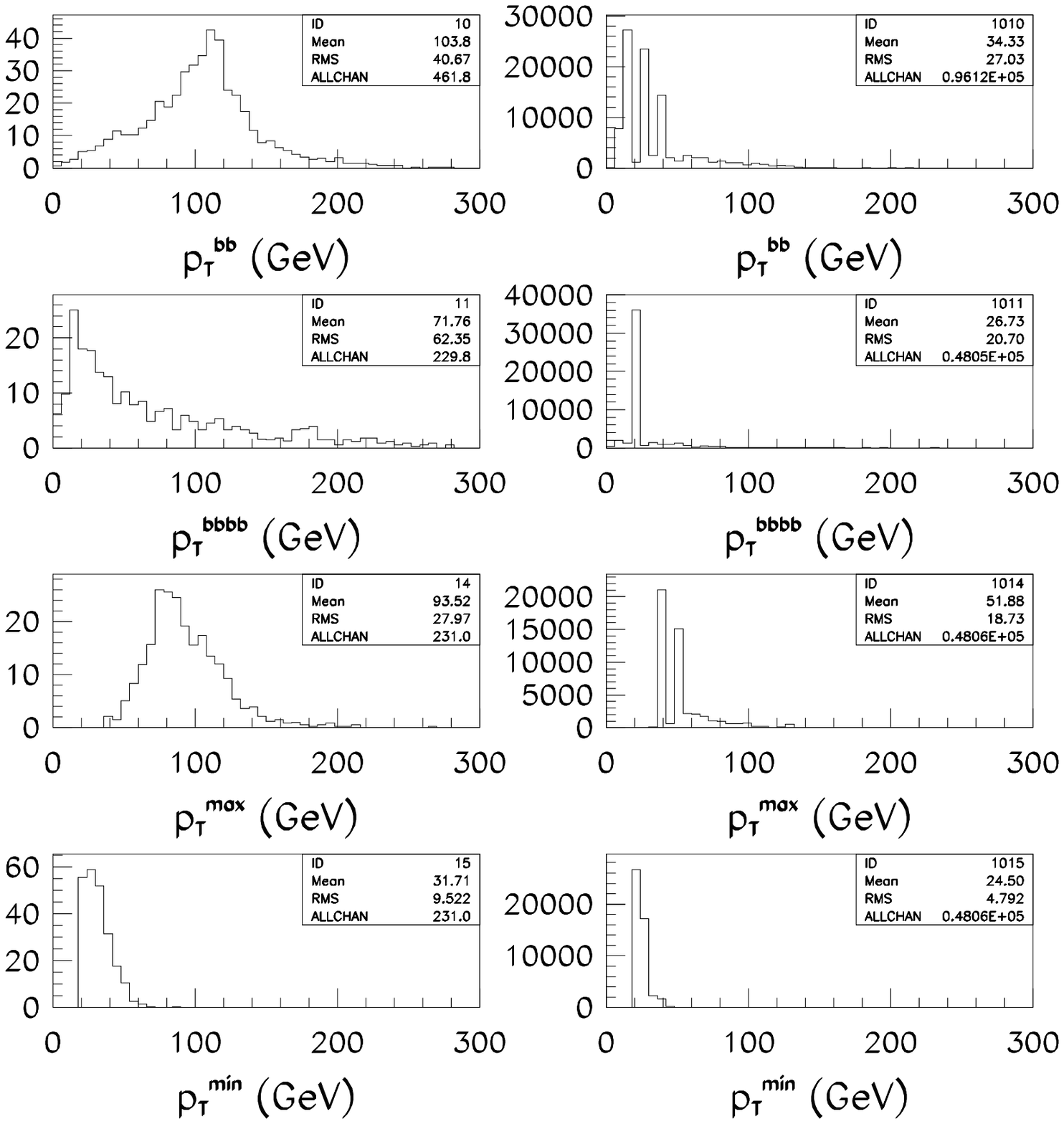,height=16cm,width=10cm}
\end{center}
\caption{\em 
 Distributions of $\mathrm{p_T^{b \bar b}}$, $\mathrm{p_T^{b \bar b b \bar b}}$,
 $\mathrm{p_T^{b_1}}$ and $\mathrm{p_T^{b_4}}$
 for signal (left) and background (right) after applying the selection B with 
 a jet $\mathrm{E_T}$-threshold of  20 GeV, for an integrated luminosity of \luma\ (ATLAS).
\label{FHIGGS:distriblow}}
\end{Fighere}

\newpage
\boldmath
\section{Conclusions}
\unboldmath

This first study of event topologies containing four b-jets in the search for  MSSM
 Higgs-boson decays has shown that it will be very difficult to cleanly extract the 
signal above
the background over a significant region of the MSSM parameter space. The results
 of this note are much less optimistic than those from \cite{Gunion}, because the study
reported in  \cite{Gunion} did not consider all processes leading to multiple b-jets
in the final state, nor did it consider the significant fraction of fake b-jets to be 
expected in such complex topologies. The study reported here was particularly difficult
 to bring to a conclusion since more than $2 \cdot 10^7$ background events had to be 
generated and processed through ATLFAST. 

The  $\mathrm{b \bar b H}$, $\mathrm{b \bar b A}$ with $\mathrm{H, A \to b \bar b}$ 
channel was considered for large
values of \mH, \mA\ and
$\mathrm{\tan \beta}$. The strongly enhanced production cross-section of 3.5 pb for \mH=500~GeV
and $\mathrm{\tan \beta} =30$ yields $10^6$ signal events for an integrated luminosity
 of \lumc. This large rate is however strongly reduced by the selection cuts
 and the b-tagging efficiencies, 
leading to roughly 5000 observable  events in a  mass bin  of $\pm 80$~GeV around 
\mH, \mA.
The QCD multi-jet background is initially several orders of magnitude higher and after
the selection cuts and the 
b-tagging procedure a  signal-to-background ratio of only a few
 percent can be achieved.
The sensitivity to this channel is therefore rather weak.

The  $\mathrm{H \to hh \to b \bar b b \bar b}$ channel was studied for low $\mathrm{\tan \beta}$ and 
\mH~=~300 GeV. Fully reconstructed $\mathrm{h \to b \bar b}$ and $\mathrm{H \to hh}$ decays in 
relatively narrow mass windows would give firm evidence for both h and H Higgs boson
production. But in the presence of the huge QCD background, the sensitivity to this channel is
also weak, weaker than that of the $\mathrm{H \to hh \to bb \gamma \gamma}$ channel.

In both cases, at most only a fraction of the MSSM parameter space, which is already 
covered by other decay modes ($\mathrm{b \bar b H, b \bar b A}$, with  $\mathrm{H, A \to \tau \tau}$ 
and $\mathrm{H \to hh \to b \bar b \gamma \gamma})$
is accessible to these channels containing four b-jets in the  final state.
Nevertheless, their observation would help in constraining
the couplings and branching ratios of the MSSM Higgs bosons.



\end{document}